\documentstyle[12pt,aaspp4,flushrt]{article}
\makeatletter

\newlength{\bibhang}
\let\@internalcite\cite
\def\cite{\let\@citeleft(\let\@citeright)%
    \@ifstar{\citeyear}{\citefull}}
\def\citenp{\let\@citeleft\relax\let\@citeright\relax
    \@ifstar{\citeyear}{\citefull}}
\def\citefull{\def\astroncite##1##2{##1~##2}\@internalcite}
\def\citeyear{\def\astroncite##1##2{##2}\@internalcite}

\def\@citex[#1]#2{\if@filesw\immediate\write\@auxout{\string\citation{#2}}\fi
  \def\@citea{}\@cite{\@for\@citeb:=#2\do
    {\@citea\def\@citea{; }\@ifundefined
       {b@\@citeb}{{\bf ?}\@warning
       {Citation `\@citeb' on page \thepage \space undefined}}%
{\csname b@\@citeb\endcsname}}}{#1}}

\def\@cite#1#2{\@citeleft#1\if@tempswa , #2\fi\@citeright}
\def\@biblabel#1{}

\makeatother

\newcommand{\PSbox}[3]{\mbox{\rule{0in}{#3}\includegraphics{#1}\hspace{#2}}}
\newcommand{\FigNum}[1]{\unitlength 1pt \begin{picture}(55,10)(-400,35) 
                        \put(0,0){Figure #1}
                        \end{picture}}
 
\newcommand{\persec}{\mbox{$\second^{-1}$}}
\newcommand{\percm}{\mbox{$\cm^{-2}$}}
\newcommand{\ppm}{\mbox{$\pm$}}
\newcommand{\cgsflux}{\erg\percm\persec}

\newcommand{\approxlt}{\mbox{$\lesssim$}}
\newcommand{\approxgt}{\mbox{$\gtrsim$}}
\newcommand{\pid}{\mbox{$P_{\rm id}$}}
\newcommand{\lr}{\mbox{$LR$}}
\newcommand{\chandra}{{\it Chandra}}
\newcommand{\usno}{{USNO~A-2}}
\def\etal{{et~al.}}

\newcommand{\tee}[1]{\mbox{$\times 10^{#1}$}}

\newcommand{\cm}{\mbox{$\rm\,cm$}}

\newcommand{\second}{\mbox{$\rm\,s$}}

\newcommand{\erg}{\mbox{$\rm\,erg$}}

\newcommand{\nbkg}{\mbox{${\rm N_{\rm bkg}}$}}

\newcommand{\voges}{V99}

\slugcomment{\it ApJ Supplements, Accepted}
\received{December 8, 1999}
\revised{March 14, 2000}
\accepted{March 31, 2000}

\begin{document}
\title{XID: Cross-Association of ROSAT/Bright Source Catalog
X-ray Sources with USNO A2 Optical Point Sources}
\author{Robert E. Rutledge, Robert J. Brunner, and Thomas A. Prince}
\affil{Division of Physics, Mathematics and Astronomy}
\affil{MS 220-47, California Institute of Technology, Pasadena, CA
91107\\ {\em rutledge@srl.caltech.edu, rb@astro.caltech.edu, prince@srl.caltech.edu} }
\author{Carol Lonsdale}
\affil{Infrared Processing and Analysis Center, Caltech/JPL, Pasadena,
CA 91125\\ {\em cjl@ipac.caltech.edu}}

\begin{abstract}

We quantitatively cross-associate the 18811 ROSAT Bright Source
Catalog (RASS/BSC) X-ray sources with optical sources in the \usno\
catalog, calculating the the probability of unique association (\pid)
between each candidate within 75\arcsec\ of the X-ray source position,
on the basis of optical magnitude and proximity.  We present catalogs
of RASS/BSC sources for which \pid$>$98\%, \pid$>$90\%, and
\pid$>$50\%, which contain 2705, 5492, and 11301 unique \usno\ optical
counterparts respectively down to the stated level of
significance. Together with identifications of objects not cataloged
in \usno\ due to their high surface brightness (M31, M32, ...) and
optical pairs, we produced a total of 11803 associations to a
probability of \pid$>$50\%.  We include in this catalog a list of
objects in the SIMBAD database within 10\arcsec\ of the \usno\
position, as an aid to identification and source classification.  This
is the first RASS/BSC counterpart catalog which provides a probability
of association between each X-ray source and counterpart, quantifying
the certainty of each individual association.  The catalog is more
useful than previous catalogs which either rely on plausibility
arguments for association, or do not aid in selecting a counterpart
between multiple off-band sources in the field.  Sources of high
probability of association can be separated out, to produce
high-quality lists of classes (Seyfert 1/2s, QSOs, RS CVns) desired
for targeted study, or for discovering new examples of known classes
(or new classes altogether) through the spectroscopic classification
of securely identified but unclassified \usno\ counterparts.  Low
\pid\ associations can be used for statistical studies and follow-on
investigation -- for example, performing follow-up spectroscopy of the
many low-mass stars to search for signatures of coronal emission, or
to investigate the relationship between X-ray emission and classes of
sources not previously well-studied for their X-ray emissions (such as
pulsating variable stars).

We find that a fraction $\sim$65.8\% of RASS/BSC sources have an
identifiable optical counterpart, down to the magnitude limit of the
\usno\ catalog which could be identified by their spatial proximity
and high optical brightness.

\end{abstract}


\section{Introduction}

The ROSAT Bright Source Catalog (BSC; \citenp{voges96,voges99})
contains positions, X-ray countrates, and spectral information of
18811 X-ray sources with countrates $>$0.05 c/s, observed during the
ROSAT All-Sky-Survey (RASS).

Efforts to identify the sources of X-ray emission with counterparts in
other wavebands are hampered by source confusion. The error-region of
the RASS/BSC sources average $\sim$12\arcsec (1$\sigma$), which can
contain several candidate objects, any of which may be the source of
X-rays.

To date, most efforts to identify the X-ray sources with parent
populations -- usually, optical sources -- have been targeted toward
sub-groups of known X-ray emitting populations, such as coronal X-ray
sources \cite{berghoefer97,huensch98a,huensch98b,huensch99}; AGN/QSOs
\cite{thomas98,beuermann99}; OB stars
\cite{berghoefer96,berghoefer96er,berghoefer97,motch97a,motch97b}; and
high galactic latitude spectrally soft sources \cite{thomas98}.  As
part of a larger effort to identify QSOs, a general spectroscopic
survey has also identified stellar type-sources \cite{bade95,bade98}.

Many of these associations -- though not all -- have been based upon
an argument of plausibility.  In the plausibility method, one
typically performs imaging photometry and spectroscopy of objects
within the X-ray error-box, and finds a plausible counterpart among
these; a counterpart is usually considered plausible if the candidate
object's class is previously known to emit X-rays, and if the
properties (such as magnitude, or implied $L_{\rm X}/L_{\rm opt}$) are
consistent with those previously observed from other objects within
its class.  This method is useful when the parent population is known
and rare (much less than one object per average X-ray error-box size).
However, this method will not discover X-ray sources independent of
object classification. In addition, some of the studies which rely
upon plausibility do not measure the level of background
contamination, while others do not evaluate the limiting \pid\ (the
lowest probability of unique identification a prospective counterpart
can have, and still be included in the catalog).  None provide a \pid\
for each cross-identification, which makes it impossible to
quantitatively evaluate the quality of a purported association in
future work.

In Table~\ref{tab:prevwork}, we list previous works which catalog
$\approxgt$ 100 optical counterparts to RASS/BSC objects, or which
sought RASS/BSC counterparts for a particular class of sources.  The
table includes: (1) the reference; (2) brief description of the
cross-identification catalog; (3) the number of cross-identifications
found and the estimated number of mis-identified (background) sources
in the cross-id list ($N_{\rm bkg}$); (4) a brief description of the
identification algorithm used; (5) and the probability of unique
association (\pid) between the X-ray source and candidate counterpart at
the identification limit of the catalog.

For several of these works, no estimation of the probability of
cross-identified sources being background sources at the detection
limit was given. Of those which do, a probability of confusion with
background (that is, unassociated) sources of $\sim$50\% is a common
limit (below which, the identified counterpart is more likely to be an
unrelated background object than actually associated with the X-ray
source.  An extensive comparison with several published and
unpublished cross-identification catalogs was made in Table~3 of the
RASS/BSC work \cite[\voges\ hereafter]{voges99} (for a total of
$\sim$17,000 sources), which we discuss more completely in
Sec.~\ref{sec:v99}.  We include, for comparison, the results of the
present work.

We have undertaken a project of off-band identification of ROSAT/BSC
X-ray sources -- XID.  The goal of this project is to provide a
catalog of cross-identifications, which provides the probability of
unique identification (\pid) between the off-band counterpart for all
RASS/BSC sources.

We have additional motivation for performing the present work.  Future
databases of sources -- both in the X-ray and in other bands -- will
and already do contain $10^{5}-10^{9}$ objects.  This is too great a
number of objects on which to perform non-automated methods for
identifying counterparts.  We therefore develop and use an automated
method for identifying cross-band counterparts, which can be further
adapted and used in future studies.

We present the method and results of a statistical
cross-identification between RASS/BSC and \usno\ catalogs, producing a
high, medium, and low-quality cross-identification list.  We summarize
these results, and provide a short discussion on the content of the
cross-identification catalog.  These catalogs are given in the
appendix. 

\section{Method}

\subsection{Data Selection}

We began with the full ROSAT Bright Source Catalog (\voges), of 18811
X-ray sources.  We use the source name, position (RA and Dec.), and
positional uncertainty.  For 57 of these sources, the catalog
positional uncertainty was given as 0\arcsec, for various reasons
specific to the source detection algorithms; we adopt, for these
objects, positional uncertainties of 30\arcsec. This makes up our
X-ray source list.

We extracted from the \usno\
catalog\footnote{http://ftp.nofs.navy.mil/projects/pmm/a2.html}, the
positions and $B$ magnitudes of all those sources which were within
75\arcsec\ of the X-ray source positions.  These sources make up the
cross-identification candidate list, the contents of which are: (1)
the name of the RASS/BSC source; (2) the positional uncertainty of the
RASS/BSC source in arcsec; (3) the name of the candidate \usno\
object; (4) the distance between the RASS/BSC and \usno\ source positions in arcsec;
(5) and the $B$ magnitude of the \usno\ object.  

We also extracted from the \usno, for each X-ray source field, ten
background fields, each 75\arcsec\ in radius.  These background fields
were offset in RA from the X-ray source position, by
$n\times$150\arcsec, for $n=[-5,-1]$ and $n=[1,5]$, five to the east
and five to the west.  These sources make up the background list.

In Fig.~\ref{fig:srcbkgcompa}, we compare the distributions in $B$
magnitude, between the cross-identification list and the background
list.  There is a clear excess of objects in the source fields at
$B<17$, indicating associated optical sources in this magnitude
range. There is a deficit of objects at faint magnitudes ($B>20$).  We
can think of no astronomical reason why there might be fewer faint
optical objects in our source fields, but we have found that, in
fields which contain bright objects ($B<14$), there are fewer faint
objects ($B>18$) than in fields which do not contain such bright objects; it
therefore seems likely that the deficit of faint objects in our source
fields are due to the excess of bright objects.  As our method is
tuned to select for the brightest objects in the field, this will not
affect our results.

In Fig.~\ref{fig:srcbkgcompb}, we compare the distribution between the
field center and optical objects for source fields (where the X-ray
position is the field center) and background fields (where there is no
X-ray source at the field center.  There is an excess of objects in
X-ray source fields within 20\arcsec\ of the X-ray source position,
and no excess of sources at greater separations.  Thus, there is a
tendency for there to be bright objects in close positional
association with the X-ray sources, simply through comparison of the
optical object characteristics in source fields and background
fields. 

\subsection{Method of Calculating of the Probability of Unique Association
(\pid)}

The applied method of cross-association is based on a method described
elsewhere \cite{lonsdale96}. This method is similar to one previously
employed \cite[and references therein]{sutherland92}, but handles
systematic uncertainties in the measured source characteristics
through statistical comparison between ``on-source'' and background
fields, whereas these previous methods assumed perfect knowledge of
source characteristics.  We here repeat essential parts of this
method, expanding upon particular modifications of the original method
as used in the present application.

We begin with the catalog of $N_x$ X-ray sources.  We consider for
each X-ray source the $M$ candidate \usno\ objects in the
cross-identification catalog. For the $i$th counterpart candidate, we
calculate a likelihood ratio -- a likelihood of association between
the optical object and the X-ray source -- through a specified
likelihood ratio method (\lr), which we take to be like a product of
probabilities:

\begin{equation}
\label{eq:lr}
LR_i =  \Pi_j P_j(x_i)
\end{equation}

\noindent where $P_j$ is a normalized probability distribution of some
property $x$ of the $i$th counterpart candidate.  The functions $P_j$
can be -- in principle -- any function of the properties $x_i$;
however, the results depend intimately upon the form of the functions
chosen, so they must be considered carefully.  By selecting functions
$P_j$ which are ratios of the {\it a priori} distributions of true
counterparts to distributions of background sources, the product
$LR_i$ has a number of useful properties.  Most importantly, $LR$ will
be high (on average) for true counterparts and low (on average) for
background sources. This permits the true counterparts to cluster at
high $LR$ values, which is desirable for a reason which will become
clear.

In the present case, we adopt a Gaussian distributed positional
coincidence for sources and an expectation that the counterpart be
very bright (low magnitude), the \lr\ for the $i$th optical source is:

\begin{equation}
\label{eq:lr2}
LR_i =  \frac{\exp{-r_i^2/2\sigma_i^2}}{\sigma_i \; N(<B_i)}
\end{equation}

\noindent where $r_i$ is the distance to the \usno\ object from the
X-ray source position in arcsec, $\sigma_i$ is the uncertainty in the
source positions (which we approximate in this case as the uncertainty
in the ROSAT/BSC source position, which is 12.7\arcsec on average, and
is typically between 5-25\arcsec, Gaussian distributed ;
\citenp{voges99}), and $N(<B)$ is the absolute number of sources in
the background list with magnitude less than some value $B$, and $B_i$
is the magnitude of the $i$th source (in the present comparison, we
use the \usno\ $B$ magnitude).

The \lr\ value is calculated for objects in the source fields and the
background fields. Following this, we calculate a Reliability $R$ of identification as a
function of likelihood ratio:

\begin{equation}
\label{eq:r}
R(LR_i) = \frac{N_{\rm true}(LR_i)}{N_{\rm true}(LR_i) + N_{\rm false}(LR_i)}
\end{equation}

\noindent which is the ratio of the number of true associations to the
sum of the true and false associations, as a function of the $LR$ of
those identifications.  $R(LR_i)$ is the binomial probability that a
optical/X-ray source pair with a specific value of $LR$ is a true
association, and that the optical source is not an unrelated
background source.  This is the probability which some previous
catalogs (cf. Table~\ref{tab:prevwork}) quote as the limiting
probability.

The probability $R$ does not, however, include the probability that
another object in the field of view could instead be the counterpart
-- that is, while the probability $R$ of an X-ray/optical pair does
state the probability of association, it may well be that there are
other sources in the field of view which are just as likely to be an
association.  In other words, the X-ray/optical association is not
{\it unique}. 

We now calculate: (1) the probability that a particular X-ray source
has no associated optical source in the \usno\ catalog ($P_{\rm
no-id}$); and (2) the probability that an optical source $i$ is the
{\it unique} association with the X-ray source ($P_{\rm id, i}$).

For a particular RASS/BSC X-ray source, there will be $M$ \usno\
objects under consideration as the possible {\it unique} association,
for which we have already calculated $M$ probabilities ($R$) for
association between the X-ray source and each optical source.  We now
impose a set of $M+1$ exclusive and complete hypotheses, and calculate
probabilities of these hypotheses being true, using the probability of
X-ray/optical associations $R$.  The hypothesis are: 

\label{sec:prob}

\begin{itemize}
\item[Hyp. $i...M$] The $i$th optical source is uniquely associated
with the X-ray source, and none of the other X-ray sources are
associated ($P_{\rm id, i}$).
\item[Hyp. $M+1$] None of the $M$ optical sources are uniquely 
associated with the X-ray source ($P_{\rm no-id}$).
\end{itemize}

Since $R(LR_i)$ is the binomial probability that the $i$th object is
associated with the X-ray source, the probability that {\it none} of
the $M$ objects are associated with the X-ray source is:

\begin{equation}
\label{eq:pnoid}
P_{\rm no-id} = \frac{\Pi_{j=1}^M (1- R_j)}{S}
\end{equation}

\noindent where $S$ is a normalization, specific to each X-ray source,
which we define below.  The probability that the X-ray source, then,
has an optical counterpart in the \usno\ catalog is $1-P_{\rm no-id}$.

For each X-ray source in the source catalog, the probability of the
association with the $i$th optical source:

\begin{equation}
\label{eq:pid}
P_{\rm id, i} = \frac{\frac{R_i}{1-R_i} \Pi_{j=1}^M (1-R_j)}{S}
\end{equation}

\noindent ($S$, again, is defined below).  This is the product of the
probability ($R_i$) that the $i$th optical object is associated with
the X-ray source and is not a background source, and the probability
that all other $j$($\ne i$) optical sources in the field are {\it not}
associated with the object and are background sources.  

$S$ is a normalization specific to each X-ray source such that the
$M+1$ hypotheses form a complete set: $P_{\rm no-id} + \Sigma_{i=1}^M
P_{\rm id, i} = 1$:

\begin{equation}
\label{eq:S}
S = {\Sigma_{i=1}^M} \, \frac{R_i}{1-R_i} \Pi_{j=1}^M (1-R_j) +
\Pi_{j=1}^M (1-R_j)
\end{equation} 

$P_{\rm id, i}$, $P_{\rm no-id}$ are now defined as functions
dependent only upon values of $R(LR)$.  Since $P_{\rm id, i}$ and
$P_{\rm no-id}$ are dependent only upon the sums and products of
$R(LR)$, they too will converge as $R(LR)$ does, according to the
central limit theorem.

For each $i$th optical source, the value $P_{\rm id, i}$ is always
less than $R(LR_i)$. This is because there can be more than one (even,
many) optical source with a high value of $R$ associated with an X-ray
source, but the set of hypotheses above excludes the possibility that
more than one of these is associated with the X-ray source (for
example, binary stellar systems, galaxy clusters, or many bright
optical sources in nearby open clusters).

Finally, we can calculate the ``quality'' of an association catalog,
which we denote by $Q$:

\begin{equation}
\label{eq:q}
Q=\langle 1-P_{\rm no-id}\rangle
\end{equation}

The value of $Q$ is the fraction of the initial X-ray catalog which
has an association in the cross-ID optical catalog.  It is
dependent only on the presence of potential counterparts in the \usno\
catalog.

It is possible to formulate an approach in which more than one optical
source is associated with the X-ray source, and we apply one approach
(for binary systems) in  \S~\ref{sec:binaries}. However, this can
quickly become a (restrictively) computationally intensive problem, as
the approach requires producing source pairs, triplets, etc., and
therefore the number of combination objects grows as the combination
factor (a factorial).  For example, there can be 15 possible optical
sources in the field of an X-ray source.  To consider only a unique
identification, there are only 15 data objects (the X-ray/optical
pair).  To consider possible binary identifications, there are
$15!/(15-2)!/2!$= 105 data objects; to consider triplet
identifications, there are $15!/(15-3)!/3!$ = 455 data objects, and so
on. 

\label{sec:S}
The means of quantifying the fraction of sources which do not fit
exactly into the unique association/no association hypotheses is
through $S$ (Eq.~\ref{eq:S}).  $S$ is the sum of all probabilities in
the unique association/no association hypotheses, and therefore, the
average value of $S$ (over all X-ray sources) is the average
probability that one of the $M+1$ hypothesis is satisfied; $1-S$ is
then the probability that none of the hypotheses are satisfied.  See
\S~\ref{sec:SS} for the quantitative estimation of $\langle
1-S\rangle$.

The values and meaning of $R$ and \pid\ are different, and should be
viewed differently.  The value of $R$ is the binomial probability of
association between the X-ray source and the \usno\ object -- or of
any \usno\ object of that $B$ magnitude and the same angular distance
from the X-ray source.  It is possible that more than one \usno\
source in the same field have high reliability (e.g. $R$=0.999). This
could occur for example, if a counterpart is one star of a binary of
equal magnitude; or if the counterpart is a star in a crowded open
cluster, and in which there may be many examples of similarly
optically bright sources; or if the plate-scanning detection algorithm
finds a saturated star and does not correctly subtract all the flux,
leaving residuals in which a second ``star'' is found.  Thus, while a
high value of $R$ does mean the source is unusual in background fields
(and therefore probably associated with the X-ray source) it does not
account for confusion -- the fact that any number of sources in the
field can have similarly high $R$, due to astronomical or systematic
considerations.  By calculating \pid\ (Eq.~\ref{eq:pid}), we find the
probability that a particular object is associated, and that {\it none
of the other objects in the field} are associated.  In this way, \pid\
accounts for confusion, while $R$ does not.  Note that in all cases,
\pid$\leq R$.

\subsection{Practical Calculation of the Probabilities}

After defining the probabilities above, we now go about estimating
them.  The sum of the number of true associations and false
associations (cf. Eq.~\ref{eq:r}) is the total number of objects (per
field) in source fields,  with the value $LR$:

\begin{equation}
\label{eq:n1}
N_{\rm true}(LR) + N_{\rm false}(LR) =  N_{\rm source}(LR) 
\end{equation}

\noindent We also observe the number of false associations as the
total number of objects (per field) in background fields with the
value $LR$:

\begin{equation}
N_{\rm false}(LR) = N_{\rm background}(LR) 
\end{equation}

\noindent Using these observed quantities, we
estimate the Reliability as a function of $LR$:

\begin{equation}
\label{eq:r2}
\tilde{R}(LR) = \frac{N_{\rm source}(LR) - N_{\rm background}(LR)}{N_{\rm
source}(LR)}
\end{equation}

\noindent where we use $\tilde R(LR)$ to indicate the calculated
approximation of $R(LR)$ (Eq.~\ref{eq:r}).  In calculating
$\tilde R(LR)$ it is necessary to generate sufficient numbers of
objects so that the uncertainty in values of $N(LR)$ is small, and so
that $\tilde R(LR)$ will converge according to the central limit
theorem.  We used $LR$ bin-sizes of $\log{LR}$=0.5 (full width), in a
running sum, centered around the source $LR$ value, except at the high
and low ends of the $LR$ distribution, where we used a single $LR$ bin
between the highest value of $\log{LR}$ and $\log{LR}-0.5$. If we
found that the bin-size was insufficient to establish $\tilde R$
to better than 0.01 (assuming Gaussian counting statistics) or there
were fewer than 1000 sources (from the source field) in a bin, we
doubled the bin-size until there were sufficient numbers of objects to
meet this criterion.  Finally, $\tilde R(LR)$ should be a
monotonically decreasing function of $LR$ for low values of $LR$
(where background sources dominate).  We chose to set $\tilde
R(LR)=0$ for all values $LR<LR_0$, where $LR_0$ is the greatest value
where $\tilde R(LR)=0$.  This is equivalent to removing from
analysis the candidate associations which are considered unlikely
counterparts according to our criteria.  In the present analysis,
$\log{LR_0}=-10.31$.  

In Fig.~\ref{fig:lr}, we show the calculated $LR$ and $R$ values upon
which the identifications are made. In the top panel, there is an
excess of sources with high $LR$ in the on-source fields compared to
the background fields.  As per Eq.~\ref{eq:r2}, this excess of sources
at high $LR$ produces the high Reliability (lower panel) for these
objects (lower panel).  The excesses are {\it statistical} -- it is
the fact that there are excess in significant numbers above that
expected from background fields produces secure identifications.

We then assign the estimated $\tilde R(LR)$ to the X-ray/optical pairs of
value $LR$.  From these $\tilde R$, we estimate the probability of unique
association according to the method in \S~\ref{sec:prob}.  

To estimate the uncertainty in the resulting \pid\ (where we have
suppressed the subscript $i$ for brevity) and $P_{\rm no-id}$, we
propagate the uncertainty in the {\it background} objects, taken to be
small and Gaussian, which is a reasonable assumption under the
conditions that the $LR$ bin from which $\tilde R$ is calculated has
either $N_{\rm source}(LR) \gg N_{\rm background}(LR)$ or $N_{\rm
background} \ge 100$ which for all our bins in the present analysis is
true.  We show in our results section the estimated uncertainty in
\pid, as a function of \pid.

We applied three different probability criteria, to produce three
catalogs of different quality: \pid$\ge$98\%, 98\%$>$\pid$\ge$90\%, and
90\%$>$\pid$\ge$50\%.

\subsection{Identifying Binary Counterparts}
\label{sec:binaries}

In some cases, there may be two or more potential counterparts in the
source field, for which the calculated Reliability is high, but the
\pid\ is low.  This will occur if, for example, the counterpart is a
bright ($B=3.0$) binary. Both resolved stars could have high
Reliabilities (say, 0.9999), but the \pid\ for both would be close to
0.5; either may be the counterpart, but the algorithm above places
high $\pid$ only when the source is unlikely to be a background
object, {\it and when there are no other sources in the field which also
are unlikely to be a background object}, the latter condition being
violated in the case of a bright binary pair.

One possible, but flawed, solution to this problem is to use the
Reliability as the indicator of a counterpart.  However, this does not
take into account the likelihood of finding a bright source near
another bright source -- such as occurs in open clusters, or in
binaries.  A second flawed approach is to use the values found for $R$
to calculate the probability of finding two objects, each of high $R$,
in a single image.  However, this assumes the values of $R$ to be
independent which,  among source fields which are clusters of objects,
is not true. 

Thus, we modified the above method to apply it for the special case of
binaries. For each object, we first calculate a likelihood ratio for
each pair in the ``candidate binary counterpart'' (compare with
Eq.~\ref{eq:lr2}):

\begin{equation}
\label{eq:lr3}
LR_{i,j} =  LR_{i} \times LR_j	
\end{equation}

\noindent This is done for each paired combination of sources in the
source field, and in the background fields, after which, the method
described for single sources applies as described above, resulting in
a list of probability of identification \pid\ for each pair of objects
in the field, a probability of no-identification $P_{\rm no-id}$ for
each X-ray source (the probability of finding a binary counterparts in
the field), and a $Q$ value (Eq.~\ref{eq:q}). Finally, we excluded
from binary-identifications those RASS/BSC sources for which a
single-object identification with \pid$>$50\% was already found. 

\subsection{Assumptions Implicit in the Quantitative Cross-ID  Method}

There are a few assumptions which are implicit in the described
association method.  We describe these here.

First, the method of determining the value of $LR$ (Eq.~\ref{eq:lr})
contains all the astronomical assumptions about the nature of the
counterparts. As such, this method finds counterparts only when the
observational characteristics of these counterparts are previously
assumed.  We have assumed that the optical counterparts will be among
the brightest optical point-sources observed, and that these sources
are spatially coincident with the X-ray counterpart.  In other
applications, one might assume that a specific $f=L_x/L_{\rm opt}$
ratio would pick out the kinds of X-ray/optical sources expected, and
in that case one can fashion a $LR$ method which would produce a high
value of $LR$ near the specified value of $f$, for example:

\begin{equation}
LR_i =   \frac{\exp{(-r_i^2/2\sigma_i^2)}}{\sqrt{2\pi}\sigma_i} \frac{\exp{(\frac{-(f_i-f_0)^2}{2
\sigma_0^2})}}{\sqrt{2\pi}\sigma_0}
\end{equation}

\noindent where $f_i$ is the ratio $ L_x/L_{\rm opt}$ of the $i$th
object, $f_0$ is an average ratio, and $\sigma_0$ is related to the
width in $f$ observed sources are expected to display, and all other
values are as defined for Eq.~\ref{eq:lr2}.  In the method we define
in Eq.~\ref{eq:lr2}. we are therefore searching for a population of
counterparts which are both within the error-region of the X-ray
source, and which have high $B$-band fluxes.  This will, quite
naturally and as we intend, find a particular class of counterparts,
which will therefore have these properties of spatial coincidence and
optical brightness.  Other classes of counterparts can be found with
different definitions of \lr.

Second, the method implicitly assumes that the properties of
background objects in the source field -- the brightness distribution
and source density -- are identical to those of the background
field. This is not necessarily the case, and can affect the calculated
probabilities.  For example, X-ray sources such as Young Stellar
Objects are often found (as we do here) in open clusters of angular
size comparable to or smaller than a RASS/BSC error region. Open
Clusters have higher source densities than an average background
field.  A higher number of unrelated objects in a source field
produces more sources with higher $R$, which will decrease the \pid\
of a particular counterpart (cf. Eq~\ref{eq:pid}).  We have attempted
to overcome this problem in the case of binary counterparts, by adding
an analysis which will find pairs of objects; this does not address,
however, counterparts with several objects in the field, such as open
clusters or galaxy clusters.  The means of completely overcoming this
downward-bias on \pid\ is superior X-ray localizations, which we
cannot obtain for a one-of-a-kind catalog such as the RASS/BSC.  Thus,
this bias exists in the catalogs we present.

Third, the method demonstrates an {\it association} between an X-ray
source and the optical counterpart -- but it does not demonstrate
unique {\it identification}. For example, if an X-ray source happens
to be an optically faint galaxy in a rich galaxy cluster of an angular
size comparable to or less than a RASS/BSC error region, this method
will likely pick out the brightest galaxy in the cluster as the
associated object (if, indeed, the galaxy is bright enough to warrant
such association), whereas the brightest galaxy is not the X-ray
emitter at all. It is, however, associated with the X-ray emitter
through their joint association with the cluster.  Another example:
the X-ray source may be hot X-ray cluster gas, which is not observed
optically (at least, not in \usno ); again, this method will pick out
the brightest galaxy in the cluster as the associated object, when it
is not the X-ray source at all.  Thus, systematic biases of the type
where optically bright sources tend to cluster with (optically faint)
X-ray sources, producing a -- by our analysis -- statistically
significant association may very well exist in our catalog; indeed,
given the known types of X-ray sources, it seems likely that they do
exist in our catalog.  These types of biases must be considered when
interpreting the values of \pid\ assigned to an association.  As an
aid to evaluating these types of biases, we have included all objects
in the SIMBAD databases which are within 10\arcsec\ of the identified
\usno\ counterpart.

\section{Results}

There were 18754 RASS/BSC objects which had 321144 possible
counterparts in a total of 24.8944 sq deg, for a source density of
12900\ppm23 source/sq deg in the counterpart catalog.  

\label{sec:blank}
There were 57 RASS/BSC objects for which no optical sources were found
in the \usno\ catalog within 75\arcsec\ of the X-ray source position.
We visually inspected the DSS
survey\footnote{http://archive.stsci.edu/dss/} images at these
locations, and found that in 30 cases, the field contained a very
bright, extended object (galaxies, globular clusters, or saturated
star), such as M31, M82, M27, M63, or M60.  Almost certainly these
regions were excluded from \usno\ scanning due to their extended, high
surface brightness emission. Although we have not quantitatively
calculated their association probability, we confidently identify them
as counterparts to the X-ray sources, estimating \pid=0.9998. (We
estimate this probability, assuming 1000 such objects in the sky, with
average radii of 3\arcmin, which thus covers 0.02\% of the total
sky). For the number of RASS/BSC objects, we expect a background
contamination of $\sim$1.  These objects are listed separately, in
Table~\ref{tab:galaxies}, along with their identification.

In other fields, there are high surface-brightness regions, likely due
to nebulae or perhaps plate defects; there are some fields which
appear to contain many point sources, which we would think \usno\
scanning should have separately found; and, there are some regions
where there are clearly no detected optical point sources at all.
These last make attractive fields for further study, to find optically
faint/X-ray bright sources such as isolated neutron stars, distant
quasars, or field LMXBs.  We compile a listing of these remaining
fields in Table~\ref{tab:none}.  In two of these fields, no background
objects were found either, although the DSS reveals a large number of
suitable optical counterpart candidates, likely implying that these
fields are not included in the \usno\ catalog. 

A total of 184446 background fields were searched, finding 3011309
objects in the background catalog, in 244.8373 sq degrees, for a
source density of 12299\ppm7 objects per sq degree.  This makes for a
surplus of $(12900\pm23 - 12299\pm7)*24.8944= 15000 \pm 600$
optically associated objects in the cross identification catalog, on
the basis of field density alone.  Since some X-ray sources are bright
optical binaries, or young-stellar-objects associated with open
clusters, or galaxy clusters, some of these excess objects are likely
to be due to higher than average field densities, though it is
difficult to estimate the magnitude of this effect.

In Fig.~\ref{fig:pnoid}, we show the cumulative distribution of the
X-ray source probability of identification (Eq.~\ref{eq:pnoid}).  This
is the probability of an individual X-ray source to have an optical
counterpart among the several \usno\ objects within its field.  There
are 1184 RASS/BSC sources ($\sim$6\%) which have at least one \usno\
source in the field, but which is either too faint or to distant to be
considered a possible counterpart in this analysis.  Approximately
$\sim$39\% of the RASS/BSC sources have a probability of
identification $>$90\%, and $\sim$67\% have better than a 50\%
probability of having an optical association within \usno.

For comparison purposes, we performed the above analysis on 10\% of
our background fields, using them as ``source fields'' and comparing
these to the other 90\% of our background fields.  In
Fig.~\ref{fig:prob}, we show the distributions of $P_{\rm id, i}$ for
all sources within the actual source fields, compared with the $P_{\rm
id, i}$ derived from objects in the background (comparison) field. The
distributions are substantially different, with no objects found with
\pid$>$10\% in the comparison field. This is because (as expected)
there are no significant excess number of objects found with high $LR$
in the comparison fields over the number found in similar background
fields, to the limit of the precision of the number statistics
(\approxlt 6\%).  This subsequently produces values of
$R$\approxlt$(1.06-1.0)/1.06\sim6\%$, which is then the highest
possible value of \pid\ for sources in the background field,
explaining why no sources with \pid$>$10\% are found in the background
field.  This comparison clarifies that an object for which \pid=80\%
(for example) does not imply that, if we were looking {\it only} in
completely the wrong areas, there would be a 20\% chance of finding a
source with this \pid value.

The plateau in the source field distribution near $\pid$=0.5, which
drops at $\pid$=0.6 is likely due to binary sources; a consequence of
our applied method is that two very bright sources in the field, which
alone would make them a likely counterpart, together mutually exclude
each other. 

We also investigated what would occur if we used spatial correlation
alone -- ignoring the brightness distribution of sources. This was
done by altering the $LR$ equation, to include only the component
based on $r$ and $\sigma$, and performing the analysis otherwise as
described.  We find zero, zero, and 5413 sources with \pid$>$98\%,
90\%, and 50\%, respectively.  Compared with the 2705, 5492, and
11301 we find when we do take $B$ into account, this demonstrates that
a substantial improvement in the statistical certainty of the
identified counterpart is made when using more than just spatial
information.  

\subsection{The Catalogs: \pid$\ge$98\% Catalog, 90\% Supplementary Catalog, and 50\%
Supplementary Catalog}
\label{sec:single}

We summarize in Table~\ref{tab:prevwork}, along with the results of
previous studies, the number of cross-identified objects in each of
the 3 cumulative catalogs, the estimated number of mis-identified
objects, and the probability of unique association at the limit of
each catalog.

We find 2705 single \usno\ objects with \pid$\ge$98\%, for an
identification rate of 14.4\%. Based on the probability of
identification for these sources, we expect a total of \nbkg$\sim$18
(0.7\%) are mis-identified as associated with the X-ray sources.  The
number of misidentified sources is found:

\begin{equation}
N_{\rm bkg} = N-\Sigma_i P_{{\rm id}, i}. 
\end{equation}

We searched the SIMBAD database for objects within 10\arcsec\ of the
identified \usno\ counterpart, for possible identification of these
optical objects and to obtain information about the environment of the
cross-identified \usno\ source; we provide the results of this search
in the catalog tables.  The 10\arcsec\ radius was chosen to account
for (some) proper motions of stars, and for astrometric uncertainty.
This will account for stars with proper motion $<$0.274\arcsec/yr
comparing observational epochs 1955.0 (for the POSS~I sources at
declinations above $-17^\circ$, as included in \usno) and 1991.5.

We systematically excluded from this list the 1RXS sources themselves,
although we note when a SIMBAD-listed object was also listed as the
1RXS source.  These lists often include objects which are likely not
the X-ray sources themselves (such as HII regions); however, including
them may help elucidate the nature of the identified \usno\
counterpart. 

In the supplementary catalog of sources with $98\%>$\pid$\ge$90\% (the
90\% Supplementary Catalog), we find an additional 2787 single
sources, for a total \pid$\ge90$\% identification rate of 29.2\%.
Based on the probability of identification for these sources, a total
of \nbkg$\sim$137 (5.0\%) of the supplementary catalog are mis-identified as the
counterpart to the X-ray sources; and a total of 155 (2.8\%) of the
combined 98\% plus 90\% Supplementary catalogs are mis-identified as
associated with the X-ray sources.  This Catalog, plus the \pid$>$98\%
Catalog, forms the \pid$\ge$90\% Catalog.

Finally, in the supplementary catalog of sources with
$90\%>$\pid$\ge$50\% (the 50\% Supplementary Catalog), we find an
additional 5809 single sources, for a total \pid$>$50\% identification
rate of 60.0\%.  Based on the probability of identification for these
sources, we expect a total of \nbkg$\sim$1879 (32\%) of the supplementary catalog
are mis-identified as the counterpart to the X-ray source.  This
Catalog, plus the 90\% Supplementary Catalog and the \pid$>$98\%
Catalog, forms the \pid$\ge$50\% Catalog.  While the counterparts in
this catalog are of potentially useful confidence (1 out of every 2 is
the optical counterpart at the limit of the catalog, with increasing
prevalence for higher \pid), we include these sources largely for
completeness, for statistical surveys, and for comparison for future
work.  Since the \pid\ is dependent {\it only} on the proximity to the
X-ray source and optical magnitude relatively rare objects which are
identified with the \usno\ counterpart either by SIMBAD or in other
work may be considered as potential counterparts.  In this context,
``relatively rare'' means (roughly) fewer than 1 in 11301 optical
sources at the quoted \usno\ magnitude per full sky.  We do not list
those RASS/BSC sources which only have potential counterparts with
\pid$<$50\% from this analysis.

In Fig.~\ref{fig:source}, we show the distributions of \usno\ $B$ and
$r$ (RASS/BSC X-ray source--\usno\ source separation) for the
\pid$\ge$98\%, the 90\% Supplementary Catalog and 50\% Supplementary
Catalog.  The \pid$\ge$98\% sources are largely limited to saturated
and $B<12$ magnitude, while the greatest number of 90\% Supplementary
sources are between 11-14 magnitude.  In positional certainty, the
catalog of lesser likelihood has counterparts which are, on average,
more distant than the closer such counterparts; even so, $>$90\% of
the found associations are within 16\arcsec.

In Fig.~\ref{fig:errs} is the distribution of formal statistical
uncertainties in the \pid\ values, for sources with \pid$>$0.9, and
sources with $0.9>$\pid$>0.5$.  The high probability sources
(\pid$>$0.9) have an absolute uncertainty of $<$0.01 for 95\% of the
sources, and $<$0.004 for 65\% of the sources.  This means the sources
identified with \pid$\ge$0.98 are distinguished from sources of lower
\pid with approximately 0.005 resolution.  Objects with lower
significance ($0.9>$\pid$>0.5$) mostly have absolute uncertainties in
the 0.01-0.03 range.

\subsection{Binary Counterparts}
\label{sec:binary}

Before excluding RASS/BSC objects which already have counterparts in
the single-source catalogs, we found 317, 619, and 3550 binary
counterparts in \pid$\ge$98\%, 98\%$>$\pid$\ge$90\%, and 
90\%$>$\pid$\ge$50\%, respectively.  After excluding, we are left with
6, 25, and 441, respectively, for a total of 472 new associations. 

5 of 6 of the 98\% sources are listed as binaries or cluster members
in the SIMBAD database, the exception being the \usno\ identifications
of 1RXSJ041003.0+863735, which are a pair of stars of nearly equal
magnitude ($B$=11.4,11.5), separated by 5 \arcsec; the sole nearby
optical source in SIMBAD is listed as a single F5IV star (HD 22701),
with $B$=6.2.

Inspection of Digital Sky Survey\footnote{http://archive.stsci.edu/dss/}
images of a few randomly selected optical sources identified as a
binary association reveals that some do not appear as convincing
binaries at all, but may have been split into two by the
scanning/detection algorithm of \usno.  However, while the association
itself may not be a binary source, such objects still indicate a
significant association, at the quoted \pid\ level, as \usno\ should
contain as many such false-splits in background fields as in on-source
fields.  Thus, the lists of ``binary'' counterparts should not be
taken to imply that the identified optical binary pair are a physical
binary, or even a pair of related objects ; it only implies that
\usno\ scanning/detection algorithm split the plate-scan into two
objects, and that our method finds that the presence of these two
objects in the RASS/BSC field is statistically unlikely by serendipity
alone.  This does, however, require that \usno\ magnitudes be viewed
critically.

\subsection{Multiplet Counterparts}
\label{sec:SS}
We found an average value $\langle 1-S \rangle=0.272$ (see
\S~\ref{sec:S}), which indicates that $\sim$27.2\% of the RASS/BSC
X-ray sources do not satisfy the unique association/no unique
association hypothesis.  Alternative hypothesis to explain these
sources include multiplet (double, triple, or more) counterparts,
where more than one USNO A2 object is associated with the X-ray
source.

\subsection{What Fraction of the RASS/BSC Sources have \usno\
Counterparts?}

We find a value of $Q$=65.2\% (Eq.~\ref{eq:q}), which means that (on
average) 34.8\% of the RASS/BSC X-ray sources have no optical
counterparts in the \usno\ catalog.  This number might be affected by
our method of setting $R=0$ for all $LR<LR_0$, where $LR_0$ is the
highest $LR$ value where $R=0$. To derive an upper-limit to the
fraction of RASS/BSC X-ray sources which have counterparts in the
\usno\ catalog, we re-performed the analysis, instead setting all
values of $R=3\sigma_R$ for all $LR<LR_0$.  The exact value of
$\sigma_R$ depends on the number of objects in each \lr\ bin, but it
is in all cases $\le$0.03; from this, we place a $3\sigma$ upper limit
on the fraction of RASS/BSC X-ray sources with optical counterparts in
the \usno\ Catalog identifiable through the present method (searching
for bright, nearby sources) at $Q\le$72.2\%.

For a limit of $B=19$, with a corresponding flux of 1.6\tee{-13}
\cgsflux\ (assuming a flat spectrum; \citenp{zombeck}) for the \usno\
and assuming a value of $F_X$=5\tee{-13} \cgsflux per RASS counts
\persec (cf. \voges) at the limit of the RASS/BSC catalog, this is a
limit of $F_x/F_{\rm opt}\approx3$.  This limit is  comparable to
values obtained from AGN, galaxies, and clusters (cf. Fig. 13, \voges)
but well above those from stars.  Thus, the remaining (unidentified)
sources may well be faint extra-galactic sources.  However, another
potential population which may contribute to the unidentified sources
are isolated neutron stars (INSs), which have $F_x/F_{\rm
opt}\approx$4\tee{4}, and it remains an open question what fraction of
the RASS/BSC is composed of these objects (two RASS/BSC sources have
been identified as INSs; \citenp{walter96,haberl97}), although a
greater number of such objects was expected \cite{blaes93}.

\section{Source Classes}

In this section, we briefly discuss the various source classes found
in the SIMBAD identifications, using the \pid$>$90\% Catalog, and the
\pid$>$50\% Catalog.  In Table~\ref{tab:sourcetypes}, we list the
number of each of several source classifications listed in SIMBAD,
found within 10\arcsec\ of these \usno\ counterparts.  By far, the
greatest number of sources here are ``unclassified''; these are \usno\
objects for which there is no source listed in the SIMBAD database
within 10\arcsec\ of the \usno\ position.

\subsection{Chromospherically Active Systems: RS CVn}

In the \pid$>$90\% Catalog, there are 116 identified counterparts
which have been previously classified as RS CVns -like systems
(including GJ 501.1=RS CVn itself).  This compares to the study of
Dempsey \etal\ \cite{dempsey93a}, reporting on detections of 112 RS
CVns in the full RASS, in which the X-ray counterparts were found for
the optically-selected catalog (optical selection); whereas we find
them by searching for the bright optical counterpart (X-ray
selection).  When we include the full \pid$\ge$50\% catalog, we find
131 RS CVns.  This is a substantial fraction of the 162 RS CVns listed
in the SIMBAD database. 

\subsection{T-Tauri Stars}
 
Of 775 T-Tauri stars classified as such in the SIMBAD database, 137
are identified with \pid$>$90\% with a RASS/BSC source, and 198
identified with \pid$>$50\%.  

\subsection{AGN and QSOs}

There are a greater fraction of extra-galactic objects (AGN, Quasars,
Seyferts, and BL Lacs) in the \pid$>$50\% Catalog than the \pid$>90$\%
Catalog, likely due to the relative optical faintness of these objects
compared to galactic objects.

\subsection{Unclassified Sources in the \pid$>90$\% Catalog}

\usno\ objects which do not have an entry in SIMBAD within 10\arcsec\
(which we call ``unclassified'') make up 25\% of the $>$90\% Catalog
(1362 objects), and 41\% (4600 objects) of the $>$50\% Catalog. A spot
check of some of the brightest such objects reveals that a large
fraction of these objects are likely to be high proper motion stars,
or objects for which the astrometry and SIMBAD positions are different
by $>$10\arcsec, although some do appear to be objects which were
previously not cataloged and classified.

In Fig.~\ref{fig:unid}, we show the X-ray countrate distribution and
\usno\ $B$-magnitude distribution of these sources.  The X-ray
countrate distribution, compared with distribution of the full
RASS/BSC catalog, shows that the unclassified sources tend slightly to
be among the fainter objects, although not exclusively so.

\section{Comparison with Other Published Cross-Identification Catalogs}

The \pid$>$50\% catalog is between 2-10$\times$ greater in size of
other published catalogs with similar limiting \pid\
(Table~\ref{tab:prevwork}), with the exception of \voges, for which we
provide a more detailed comparison below.  However, we note that
previous work has used what we define as $R$ as their catalog
probability limit, whereas we use \pid, which is always less than or
equal to $R$. For example, while we find 11301 sources with
\pid$\ge$50\%, we find 12462 sources with $R\ge$50\%.

\label{sec:v99}
\voges\ presented results of cross-identifications with 16 different
catalogs of various types of sources (optical, radio).  In the largest
such comparison, they describe cross-identifications with the Hubble
Space Telescope Guide Star Catalog
\cite[HST-GSC]{lasker90,russel90,jenker90,taff90}, for which $R$=50\%
at a distance $\sim$24\arcsec\ from the RASS-BSC position than
expected from (the background) source density extrapolation from
further away (40-60\arcsec); further, this extrapolation indicated
that 13.92\% of the 15824 HST-GSC objects within 24\arcsec\ of
RASS/BSC X-ray sources were background sources, with the remainder
being associated with the RASS/BSC X-ray source. Of the HST-GSC
objects, 9759 were the sole object in the 24\arcsec\ field, making
them unique identifications down to \pid=50\%, with a contamination
rate of 13.92\%. In the remaining 6056 fields, multiple objects either
indicate an association with clustered sources (galaxies, stars), or
confusion in the true, unique association.  The HST-GSC results of
\voges\ are consistent the results of the present work (11301 objects,
to \pid=50\%, contamination rate of 18\%).

In addition, comparisons with many different cross-identification
catalogs were performed by \voges\ (NVSS, Tycho, FIRST, EUVE, and
IRAS, for example), and the primary statistical result for each
catalog was a search radius, at which $R$=50\% (W. Voges,
priv. comm.), using exclusively spatial proximity.  In these
comparisons, \voges\ found 17017 possible counterparts within
90\arcsec\ of the RASS/BSC position, 7117 of which are the sole object
in the 90\arcsec\ region.  As with the HST-GSC results, a search
radius was found for each catalog at which \pid=50\%. The number of
candidate objects within the search radius, the estimated background
source contamination, and the search radius itself varies from catalog
to catalog.

\voges\ associations are made exclusively on proximity between the
cross-id and the RASS/BSC source (the closest object is the most
likely counterpart). In contrast, the algorithm in the present work
also makes use of $B$-band brightness.  Thus, a brighter $B$-band
object which is further from the RASS/BSC source from a fainter
$B$-band object can be identified as a counterpart (if bright enough).
On the other hand, objects which have unusually {\it high} $B$
magnitude would be considered highly unlikely counterparts in the
present work (cf. Eq.~\ref{eq:lr2}, while in \voges, they may be
considered a possible counterpart on the basis of spatial proximity
alone.  In the present work, we have evaluated a unique likelihood of
association (\pid) for each object, while individual object \pid's are
not available in \voges.  Thus, while the present and \voges\ catalogs
are similar in size, future work based on the present catalog can
select out high-quality identifications for targeted work, or draw
more broadly upon the lower-quality identifications for statistical
studies based on the unique \pid\ for each counterpart.

\section{Catalog Access and Contents}

The 3 catalogs, given in the Appendix,  contain: (1) a list of the ROSAT/BSC
object by name; (2)the RASS/BSC countrate and uncertainty; (3) the
\pid\ with the identified \usno\ counterpart; (4) the \usno\ $B$
magnitude; (5) \usno\ name/position (hhmmss.ss+-ddmmss.s J2000); (6)
the name of SIMBAD objects within 10\arcsec\ of the \usno\ source; (7)
the source classification listed in SIMBAD for these objects (variable
star, binary system, galaxy, etc.); (8) the source type listed in
SIMBAD (stellar spectral type or galaxy type); (9) SIMBAD $B$ and $V$
magnitudes; and (10) accompanying source notes, including a flag if
that SIMBAD object was previously identified as the RASS/BSC source.
If there are more than one SIMBAD objects within 10\arcsec\ of the
\usno\ source, these are listed on subsequent lines.  It is not implied
that the SIMBAD objects {\it are} the \usno\ counterpart although we
expect this to often be the case, as can be told by comparing the
\usno\ $B$ magnitude to that reported by SIMBAD.  The SIMBAD objects
are listed to suggest them as the \usno\ counterpart, or to at least
potentially provide information about the counterpart's environment
(such as in a cluster of galaxies, or a stellar cluster).

\section{Discussion and Conclusions}

We have cross-correlated the 18811 RASS/BSC X-ray sources with 321144
candidate \usno\ optical counterparts within 75\arcsec\ of the
RASS/BSC source position, on the basis of $B$ magnitude and source
proximity, taking into account the quoted RASS/BSC positional
uncertainty.  On this basis, we identify 2705 \usno\ objects with
\pid$>$98\%, with $\sim$0.66\% background contamination; 5492 with
\pid $>$90\%, with $\sim$2.8\% background contamination; and 11301
with \pid$>$50\%, with $\sim18\%$ background contamination.  Thus, we have
identified possible optical counterparts to 60\% of the ROSAT/BSC on
the basis of position and photometry alone.  We have also provided --
for the first time -- a probability of unique identification between
each of the X-ray sources and their proposed counterpart.  When we
include unique ``binary'' identifications, and 30 high-surface
brightness objects which were not included in \usno, we have presented
optical associations for a total of 11803 objects, down to a limiting
identification probability of 50\%, which is 62.7\% of the RASS/BSC
catalog objects. More conservatively, we have presented optical
associations for 5553 objects, to \pid$\ge$90\%, which is 29.5\% of
the RASS/BSC catalog. The breakdown of these identified sources is
listed in Table~\ref{tab:number}.

The individual identifications are subject to systematic uncertainty
of the association between X-ray sources and clustered optical sources
(such as clusters of galaxies, open stellar clusters and star
formation regions), in which the X-ray source may reside, and the
greater than average density of candidate optical counterparts makes
the presence of a brighter-than-average source more likely than in a
background field.  Thus, the given optical identification should be
considered an ``association'' -- and the likelihood that the source of
X-ray emission is the identified optical point source directly or a
nearby associated object must be evaluated on a case-by-case basis, on
the basis of the likelihood of such a secondary association. 

For these sources, we have listed the RASS/BSC source-name, and the
identified \usno\ counterpart.  In addition, we compiled a list of
objects in the SIMBAD database within 10\arcsec\ of the \usno\
counterpart, many of which are likely to be the \usno\ counterpart
itself.  There are a surprisingly high fraction (25\% in the
\pid$>$90\% Catalog) of optical counterparts which are not named in
the SIMBAD database.  As these are (photometrically) identical to
objects which have been previously classified, the unclassified
objects are likely to be the same population.  Thus, a program of
classification of these unclassified objects will likely discover new
examples of known classes of sources, although they may contain
unknown classes as well.

The limit on the fraction of RASS/BSC sources which have counterparts
in the \usno\ catalog discoverable by this method is $Q\le$72.2\%. To
improve this identification fraction between X-ray sources and
optical data, either additional optical information is required
(source classes, spectral colors) which will help distinguish
identifiable sources, or improved X-ray localizations (such as from
the ROSAT/HRI, or \chandra), or combining X-ray and optical
information to pick out sources of particular classes ($L_x/L_{\rm
opt}$).

\acknowledgments

The authors are grateful to the anonymous referee for careful reading
of the manuscript.  This paper was produced under the Digital Sky
Project of the NPACI program (NSF Cooperative Agreement ACI-96-19020)
and NASA grant NAG5-3239.  This research has made use of the Simbad
database, operated at CDS, Strasbourg, France. This research has made
use of the NASA/IPAC Extragalactic Database (NED) which is operated by
the Jet Propulsion Laboratory, California Institute of Technology,
under contract with the National Aeronautics and Space Administration.


\clearpage

\begin{figure}[htb]
\caption{ \label{fig:srcbkgcompa} 
Comparison between cross-identification catalog objects (solid line)
and background field catalog objects (broken line). Panel  {\it
a}: Distributions of $B$ magnitude.  Panel {\it b}: Difference between
distribution in $B$ magnitude. }
\end{figure}

\begin{figure}[htb]
\caption{ \label{fig:srcbkgcompb} Comparison between
cross-identification catalog objects (solid line) and background field
catalog objects (broken line). Panel {\it a}: Distributions in
distance between X-ray source position and optical source.  Panel {\it
b}. Differences between distributions shown in panel {\it a}.  }
\end{figure}

\begin{figure}[htb]
\caption{ \label{fig:lr} Top Panel: Distribution of calculated LR
values for RASS/BSC-\usno\ candidate cross-identifications in
On-Source fields (solid line) and in background fields (broken line)
-- $n(LR)$, which is number of objects per field per LR bin.  Note the
excess of such sources at $LR>-10$, indicating optical-X-ray
associations. Bottom Panel: The Reliability -- $R(LR)$
(Eq.~\ref{eq:r2}) -- can be thought of as the probability that the
optical source under consideration is {\it not} a background source,
that it is associated with the X-ray source.}
\end{figure}

\begin{figure}[htb]
\caption{ \label{fig:pnoid} 
Cumulative distribution of the probability that the \usno\ catalog
contains a cross-identification of the 18754 RASS/BSC objects for
which at least one candidate \usno\ object was found. 
}
\end{figure}

\begin{figure}[htb]
\caption{ \label{fig:prob} Comparison between the single-source
probabilities ($P_{\rm id, i}$) in the source fields (solid line)
vs. in the background comparison fields (broken line).  This
comparison was performed to demonstrate what would happen if the
analysis were applied to fields in which are not the X-ray source
fields.  There are no significant excess optical sources found (up to
$\sim$5\%) at high LR, which limits the maximum $R$ to
$<(1.05-1.0)/1.05$, and thus \pid\ to this value as well.  The slight
excess near $p$=0.5 is possibly due to binaries in the source fields.
}
\end{figure}

\begin{figure}[htb]
\caption{ \label{fig:source} Observed $B$ (panel a) distribution and
$r$ (distance between X-ray source position and the associated
counterpart position; panel b) for the \pid$\ge$98\% catalog (solid
line), the $\ge$90\% Supplementary (dotted line) and $\ge$50\%
Supplementary Catalogs.  The 98\% sources are all $B<$12.5, the 90\%
extend down to $B$=14, and the 50\% sources all the way down to
$B=17$.  The average X-ray/optical source separations are 7.4\arcsec
(standard deviation 4.6\arcsec) and 10.3\arcsec (std dev. 7.0\arcsec)
and 11.9 (std. dev 9.2\arcsec) for the 98\% catalog, and 90\% and 50\%
Supplementary catalogs, respectively.}
\end{figure}

\begin{figure}[htb]
\caption{ \label{fig:errs} The distributions of uncertainty in \pid\
due to the uncertainty in the number of background sources, for
sources with \pid$>$0.90 (dotted line) and 0.90$>$\pid$>$0.50 (solid
line).  The small median value of these uncertainties for \pid$>$0.90
makes the distinction between \pid$>$0.98 sources and \pid$>$0.90 a
meaningful one. These uncertainties are, essentially, the resolution
of \pid\ in the indicated \pid\ ranges.}
\end{figure}

\begin{figure}[htb]
\caption{ \label{fig:unid} {\it Top Panel:} RASS/BSC countrate
distribution of un-identified sources (see text) and all RASS/BSC
objects.  There is a tendency for the un-identified sources to be
among the fainter objects, although this is not strong. {\it Bottom
Panel}: $B$ magnitude distribution of the unidentified sources.  }
\end{figure}

\clearpage
\pagestyle{empty}
\begin{figure}[htb]
\PSbox{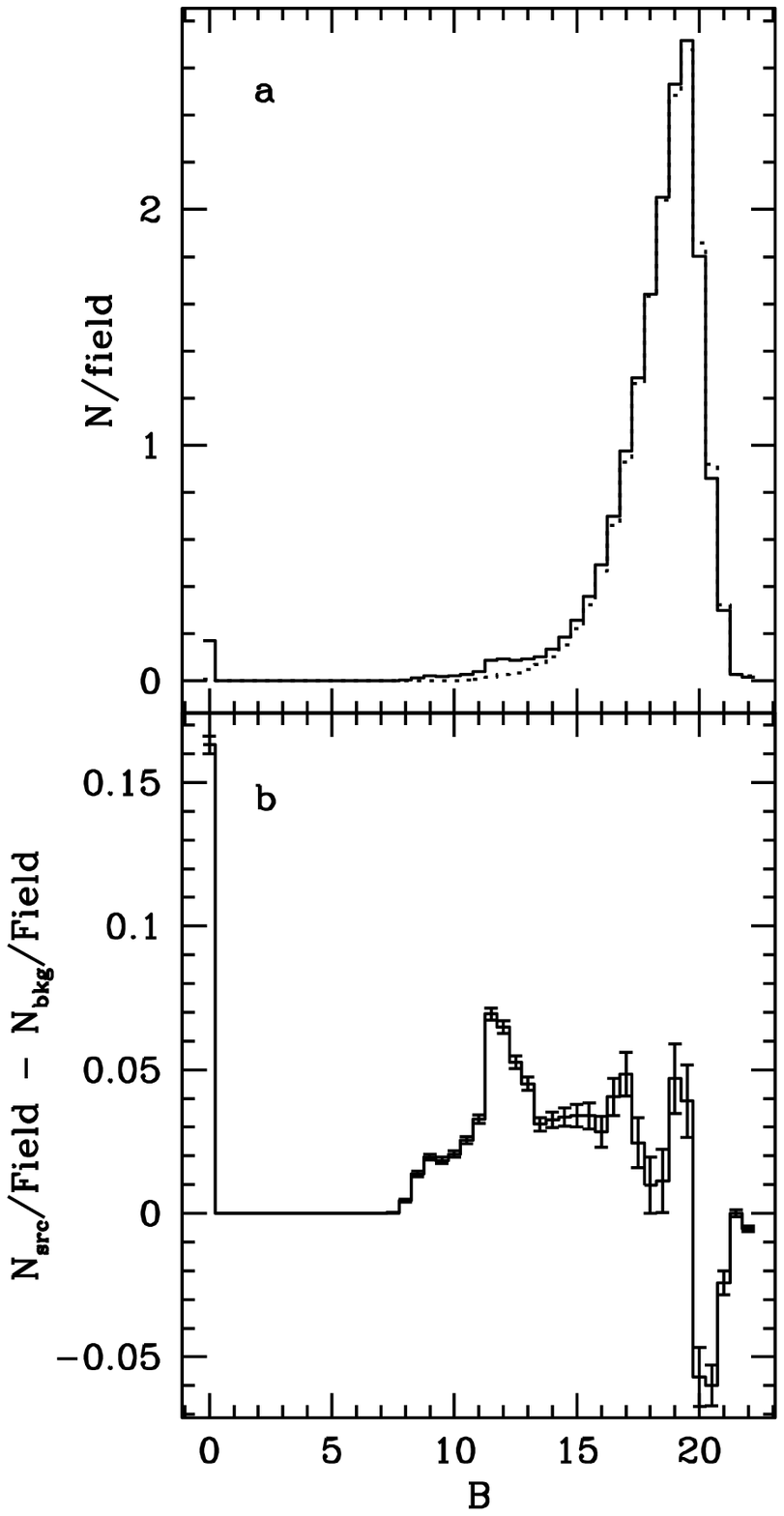 hoffset=-80 voffset=-80}{14.7cm}{21.5cm}
\FigNum{\ref{fig:srcbkgcompa}}
\end{figure}

\clearpage
\pagestyle{empty}
\begin{figure}[htb]
\PSbox{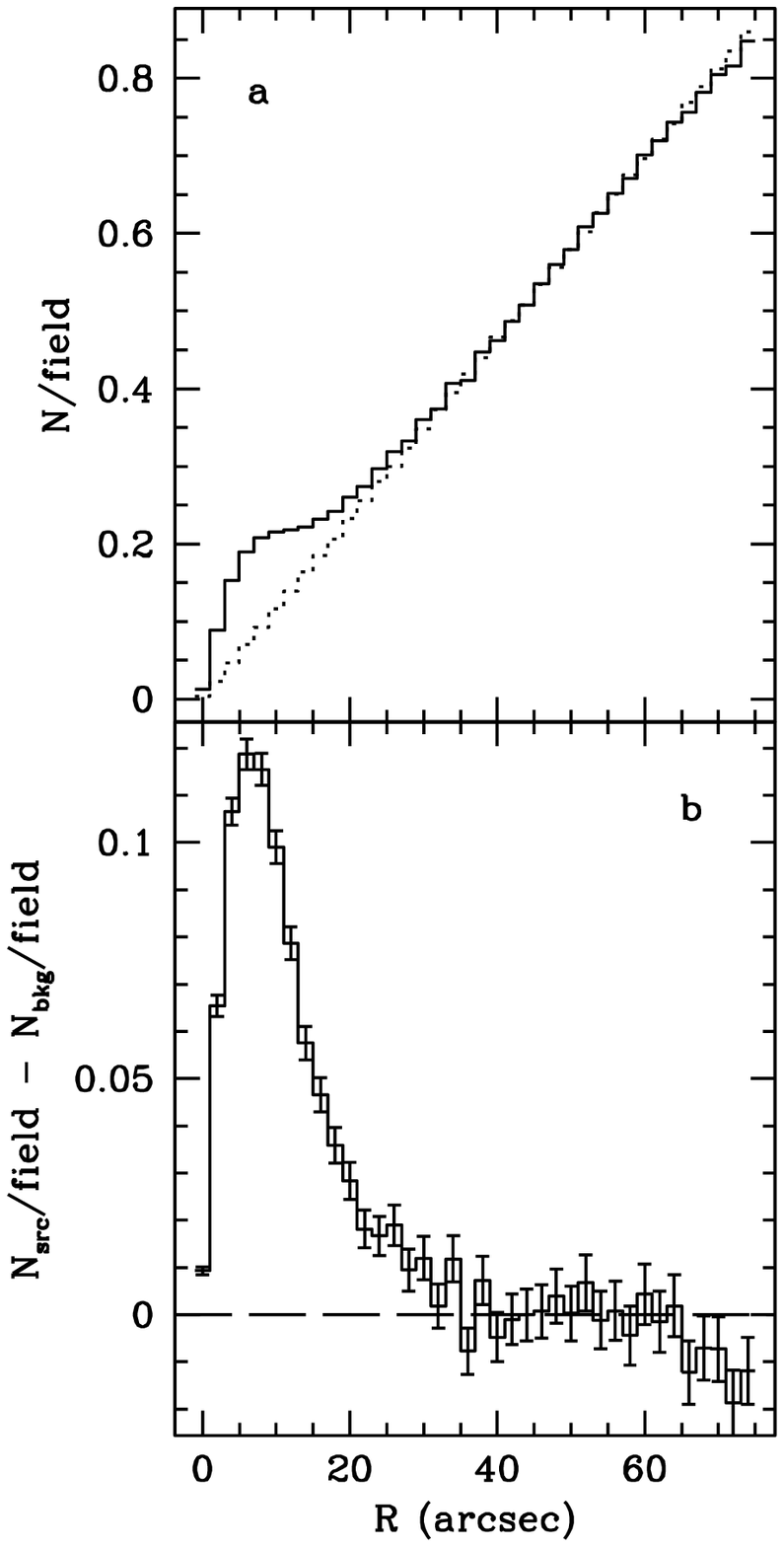 hoffset=-80 voffset=-80}{14.7cm}{21.5cm}
\FigNum{\ref{fig:srcbkgcompb}}
\end{figure}

\clearpage
\pagestyle{empty}
\begin{figure}[htb]
\PSbox{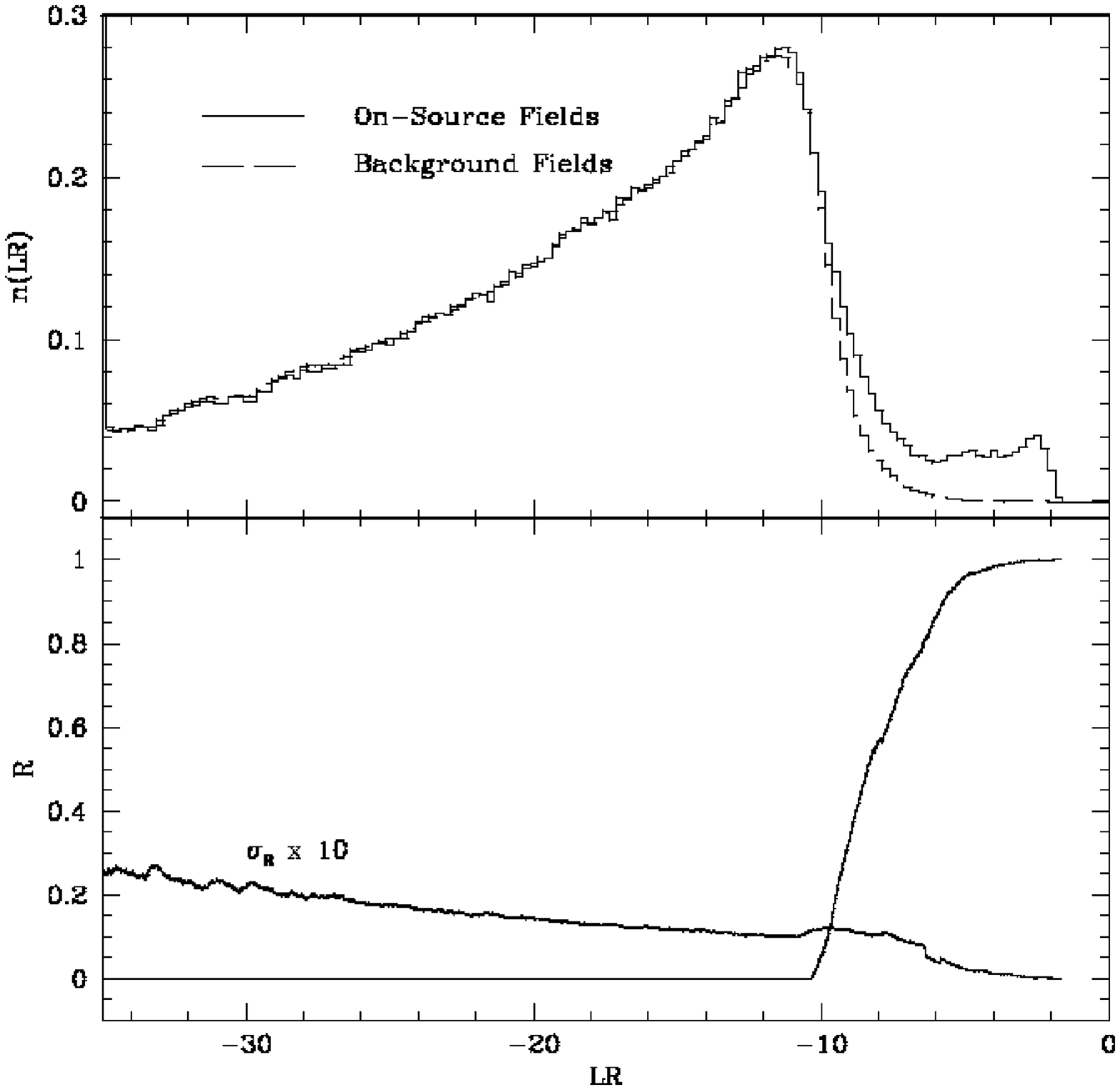 hoffset=-80 voffset=-80}{14.7cm}{21.5cm}
\FigNum{\ref{fig:lr}}
\end{figure}

\clearpage
\pagestyle{empty}
\begin{figure}[htb]
\PSbox{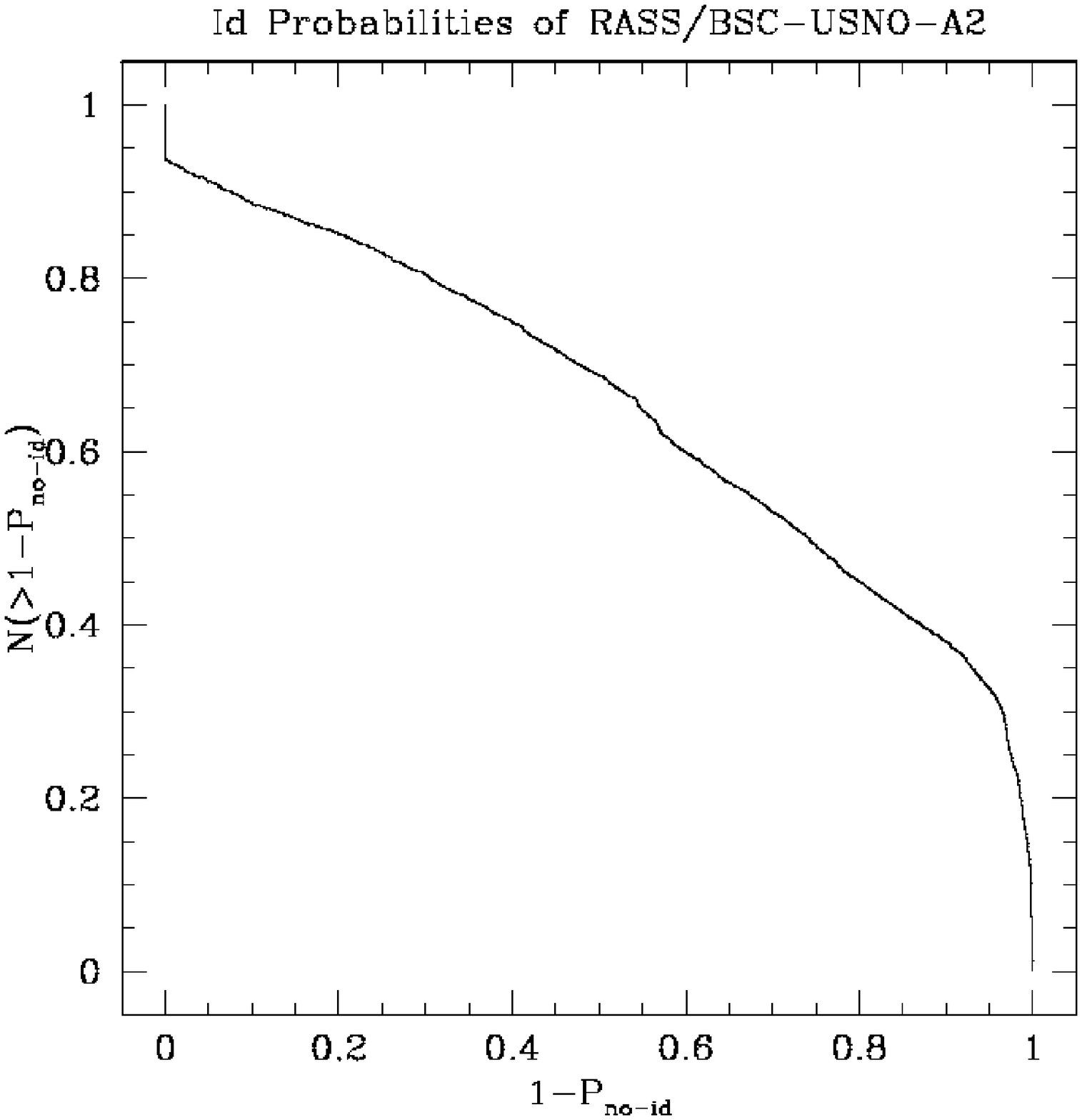 hoffset=-80 voffset=-80}{14.7cm}{21.5cm}
\FigNum{\ref{fig:pnoid}}
\end{figure}

\clearpage
\pagestyle{empty}
\begin{figure}[htb]
\PSbox{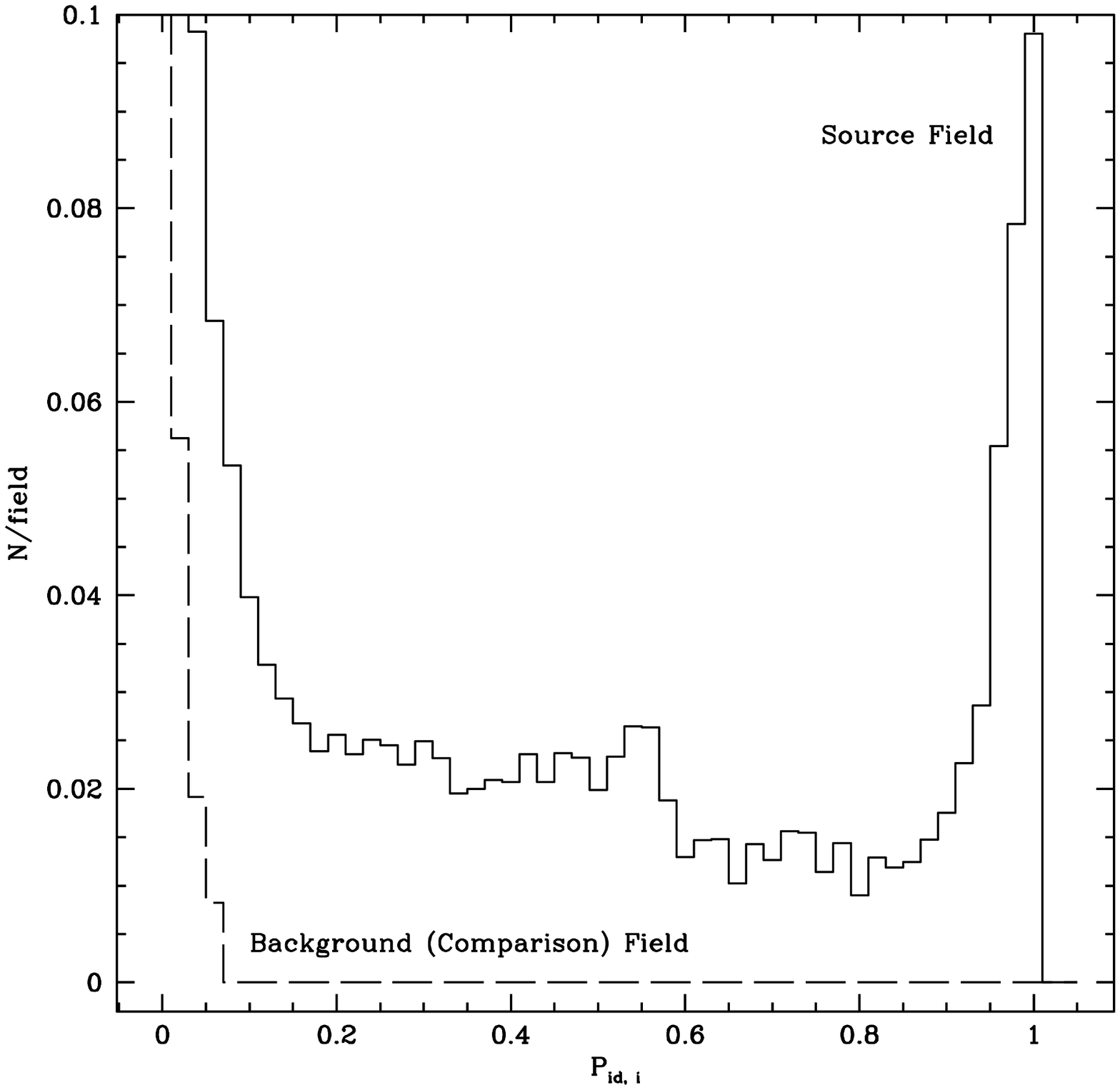 hoffset=-80 voffset=-80}{14.7cm}{21.5cm}
\FigNum{\ref{fig:prob}}
\end{figure}

\clearpage
\pagestyle{empty}
\begin{figure}[htb]
\PSbox{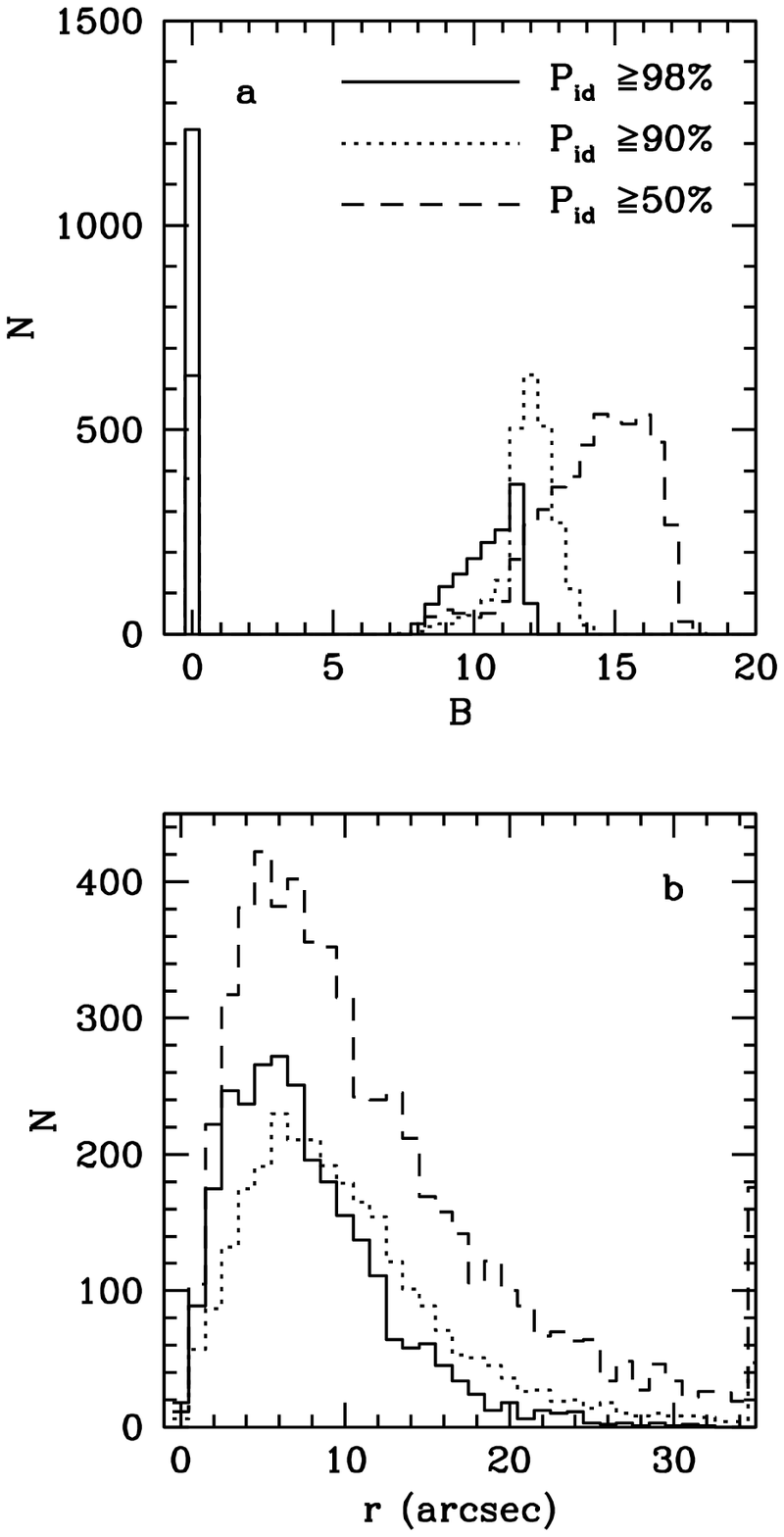 hoffset=-80 voffset=-80}{14.7cm}{21.5cm}
\FigNum{\ref{fig:source}}
\end{figure}

\clearpage
\pagestyle{empty}
\begin{figure}[htb]
\PSbox{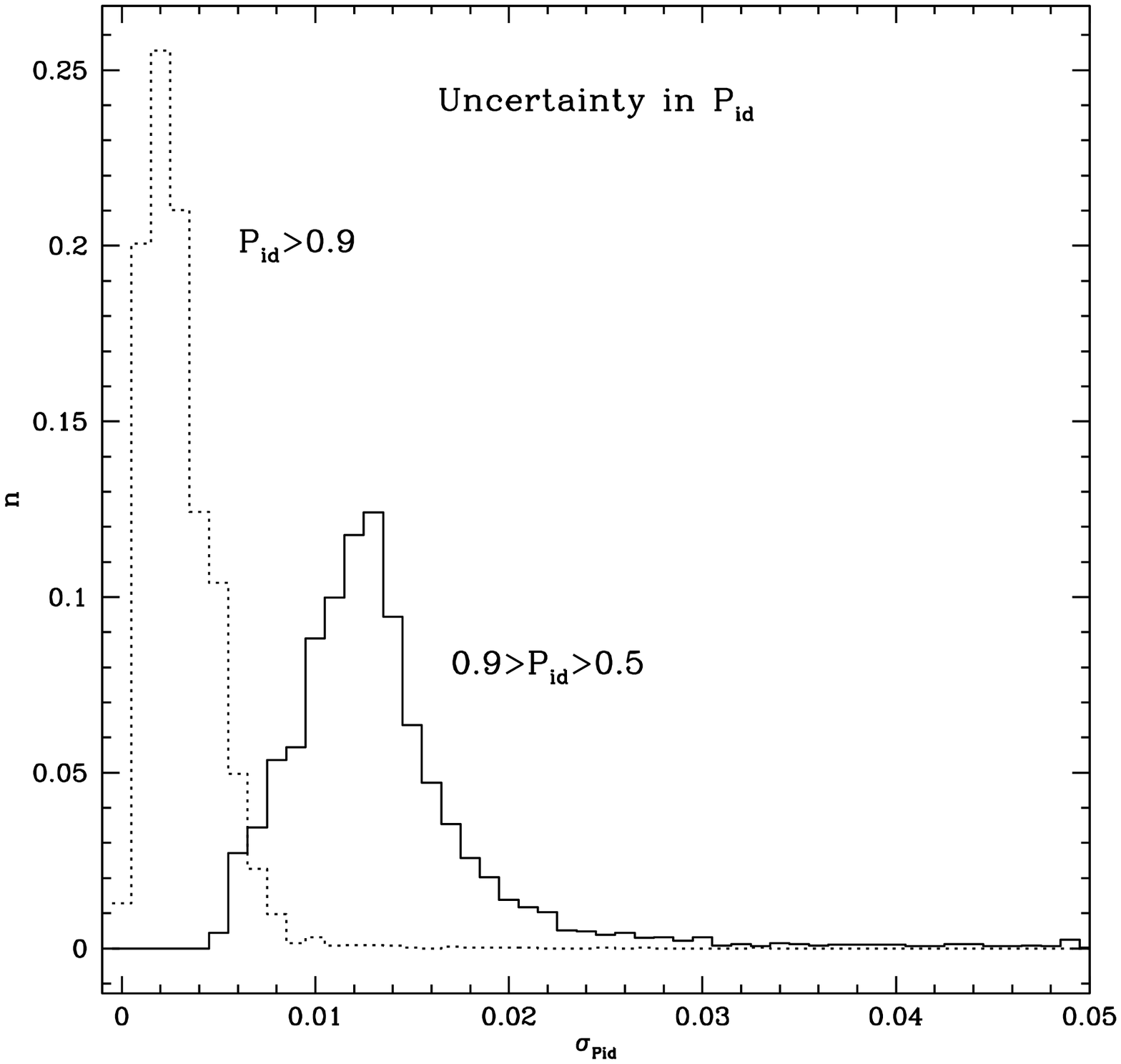 hoffset=-80 voffset=-80}{14.7cm}{21.5cm}
\FigNum{\ref{fig:unid}}
\end{figure}

\clearpage
\pagestyle{empty}
\begin{figure}[htb]
\PSbox{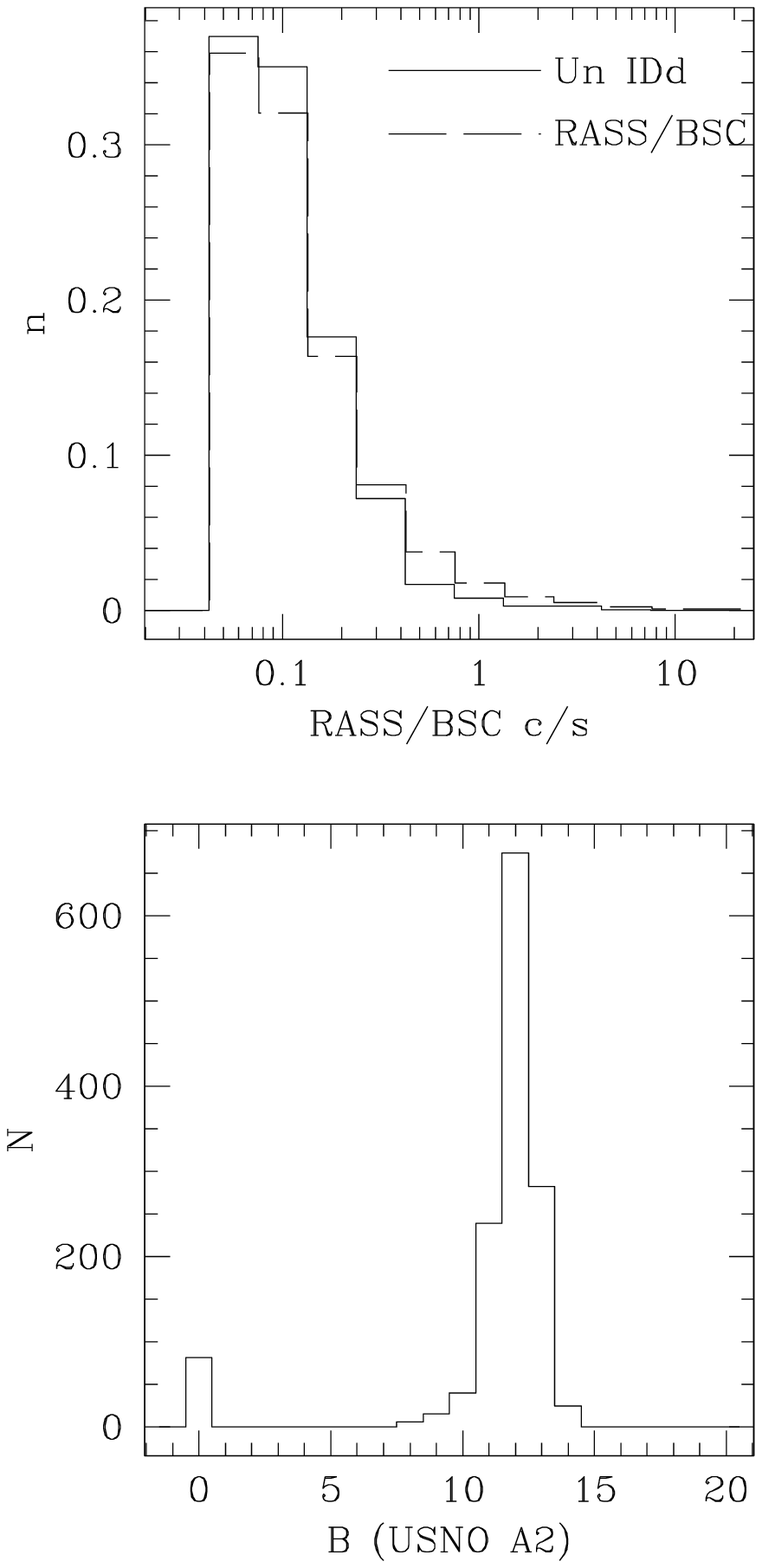 hoffset=-80 voffset=-80}{14.7cm}{21.5cm}
\FigNum{\ref{fig:unid}}
\end{figure}

\begin{deluxetable}{lllll}
\scriptsize
\tablecaption{Published RASS/BSC Counterpart Catalogs ($N\approxgt 100$)
\label{tab:prevwork}}
\tablehead{
\colhead{Ref.} & 
\colhead{Cross-ID Catalog} & 
\colhead{$N_{\rm src}$ ($N_{\rm bkg}$=\%)} & 
\colhead{Criteria for ID} & 
\colhead{Limiting Significance \tablenotemark{a}} } 
\startdata
1 & 
Stars (AFGKM; I through III-IV); & 
450 (21.8=4.8\%)	& 
$<$90\arcsec from X-ray source  &
$R>$50\% 		\\
	&
 from Yale BSC \cite{hoffleit91} & 
 & 
 & 
 \\


2 & 
Stars (AFGK class IV, V and subtypes) & 
980 (21.8=2.2\%) & 
$<$90\arcsec from X-ray source  & 
$R>$50\%  		\\
 & 
 from Yale BSC& 
 &
 &
 \\

3 &
Nearby Stars \cite{gliese91} &
1252 (24=1.9\%) &
$<$90\arcsec\ from X-ray source & 
$R>$50\% \\

4 & 
RASS/BSC: Medium-bright, &
75 (n/a) &
Plausibility &
n/a \\
 &
spectrally-soft, $|b|>40^\circ$&
 &
 &
 \\

 &
optical/spectral survey &
 &
 &
 \\

5 &
OB stars in the Yale BSC&
216 (0.5=0.23\%) & 
$<$45\arcsec\ from X-ray source & 
n/a \\  

6 &
RASS/BSC: Bright ($>0.5$ c/s), &
397 (n/a) & 
Plausibility & 
n/a \\
 &
spectrally soft (HR1$<$0), $|b|>20\deg$ &
 &
 &
 \\

7 & 
Cygnus:  $86^\circ<l<94^\circ$, $-5^\circ<b<5^\circ$ &
128 (2=1.5\%) & 
Plausibility & 
$R>$98\% \\



8 &
Full Sky, optical/spectral survey & 
3847  ($<$9=0.2\%) & 
Plausibility &
n/a \\  \tableline

9 &
HST-GSC& 
9759  (1358=13.92\%) & 
$<24$\arcsec\ from X-ray source &
$R>$50\% \\


Present work &
USNO A2, $B$ objects &
2705 (18=0.7\%) &
\pid $\ge$98\% (see text) & 
\pid$\ge$98\% \\

Present work &
USNO A2, $B$  objects & 
5492 (155=2.8\%) &
\pid $\ge$90\% (see text) &
\pid$\ge$90\% \\

Present work &
USNO A2, $B$ objects & 
11301 (2034=18\%) &
\pid $\ge$50\% (see text) &
\pid$\ge$50\% \\

\enddata
\tablenotetext{a}{At the limit of the catalog -- previous work used
$R$ (Eq.~\ref{eq:r2}), which does not include probability of source confusion;
present work uses \pid, which does include probability source
confusion (Eq.~\ref{eq:pid}) }
\tablerefs{
1: \citenp{huensch98a} ; 
2: \citenp{huensch98b} ; 
3: \citenp{huensch99} ; 
4:\citenp{beuermann99}; 
5: Berghoefer \etal 1996, 1997a,b
\nocite{berghoefer96,berghoefer96er,berghoefer97}; 
6: \citenp{thomas98}; 
7: Motch \etal\ 1997a,b \nocite{motch97a,motch97b}; 
8: Bade \etal\ 1995, 1997 \nocite{bade95,bade98}; 
9: Voges \etal\ 1999 \nocite{voges99}
}

\end{deluxetable}

\begin{deluxetable}{lrr}
\scriptsize
\tablecaption{Bright/Extended Object Counterparts \label{tab:galaxies}}
\tablehead{
\colhead{1RXS} & 
\colhead{Visual Inspection}& 
\colhead{Simbad ID}}
\startdata
1RXSJ004241.8+411535   & Galaxy &  M31				\\
1RXSJ004733.3$-$251722 & Galaxy	 & NGC 253 (G)	 \\
1RXSJ013350.9+303932   &  Galaxy  &  M 33	 \\
1RXSJ024620.0$-$301639 &  Galaxy  & NGC 1097 (Sy 2)			 \\
1RXSJ031819.4$-$662912 &  Galaxy	& NGC 1313 (G)			 \\
1RXSJ032241.8$-$371239 &  Galaxy	& NGC 1316 (GiC)		 \\
1RXSJ033828.8$-$352701 &  Galaxy	&NGC 1399	(GiC) \\
1RXSJ033851.5$-$353543 &  Galaxy	&NGC 1404	(GiC)			 \\
1RXSJ041611.4$-$554630 &  Galaxy	&	NGC 1553 (GiG)			 \\
1RXSJ042000.5$-$545617 &  Galaxy	& 	NGC 1566 (Sy1)			 \\
1RXSJ051406.6$-$400234 & Gl.  Cluster 	& 	NGC 1851	 \\
1RXSJ053803.8$-$690925 &  Galaxy	& 30 Doradus	\\
1RXSJ095534.7+690338   &  Galaxy	&  M 81				\\
1RXSJ111811.1+313154   & Bright Sat. Binary& * 53 UMa 	\\
1RXSJ112016.7+125917   &  Galaxy	& M 66	\\
1RXSJ121900.4+471747   &  Galaxy	&  M 106 (Sy 2)	\\
1RXSJ123939.6$-$052035 &  Galaxy	& NGC 4593 (Sy 1)	\\
1RXSJ124340.6+113309   &  Galaxy 	& M 60 (pair)	\\	
1RXSJ125052.5+410713   &  Galaxy		&	M 94 (LIN)		\\
1RXSJ130528.0$-$492758 &  Galaxy		&	NGC 4945 (G)			\\	
1RXSJ131549.3+420154   &  Galaxy		& 		NGC 4945 		\\	
1RXSJ132527.3$-$430105 &  Galaxy		& M 63 (G); QSO 1313+422 		\\
1RXSJ132542.9$-$425746 &  Galaxy (offset)	& M 63 		\\
1RXSJ132953.8+471143   &  Galaxy		& M 51/NGC 5194 (PoG)					\\
1RXSJ133657.0$-$295207 &  Galaxy		& 	M 83				\\
1RXSJ134210.2+282250   & Gl. Cluster	&	NGC 5272		\\
1RXSJ175012.8$-$370306 & Gl. Cluster	&	NGC 6441	\\
1RXSJ195936.2+224309   & PN 			& 	M 27 (Dumb-bell Nebula)				\\
1RXSJ212958.4+120959   & Gl.  Cluster	&	M 15 				\\
1RXSJ220916.6$-$471002 & Galaxy		& 	NGC 7213  (Sy 1)				\\

\enddata
\tablecomments{Bright or extended objects identified by visual
inspection of the DSS plate, which were originally found because there
were no \usno\ objects listed within 75\arcsec\ of the RASS/BSC
position of the X-ray source.  Estimated significance of the
cross-identification of these objects is \pid=0.999, and they are
included in the \pid$\ge$98\% Catalog.}
\end{deluxetable}

\begin{deluxetable}{lcr}
\scriptsize
\tablecaption{RASS/BSC Objects with No \usno\ objects $<$75\arcsec\ \label{tab:none}}
\tablehead{
\colhead{1RXS} & 
\colhead{N Bkg Objs.\tablenotemark{a}} &
\colhead{Visual Inspection\tablenotemark{b}}}
\startdata
1RXSJ000235.9$-$081518 & 25 	&	(no object) \\
1RXSJ002941.1$-$165408 & 26 	&	(no object)\\
1RXSJ004202.7$-$143557 & 24  	&	(no object) \\
1RXSJ040358.2$-$021113 & 29 	& (no object)	\\
1RXSJ052749.6$-$695412 & 98 	& Neb. 	\\
1RXSJ053428.3$-$052414 & 8  	& Neb. \\
1RXSJ053510.8$-$044850 & 22  	& Neb. 	\\
1RXSJ054045.7$-$021119 & 25 	& Diff. Spike	\\
1RXSJ055054.2$-$621454 & 0 	& Many Pt Src/ Galaxy?\\
1RXSJ055225.0$-$640206 & 0  	& Many Pt Src \\
1RXSJ064045.4+094927 & 46 	& Neb.		\\
1RXSJ100407.9+144925 & 30 	& (no object)	\\
1RXSJ104346.4$-$594538 & 7 	& (no object) \\
1RXSJ111005.5$-$763531 & 191 	&  Neb. 	\\
1RXSJ123607.4+731901 & 45 	& (no object)	\\
1RXSJ124601.5$-$680846 & 370 	& (no object/star?)	\\
1RXSJ124634.5$-$680446 & 373 	& (no object/star?, same as above)	\\
1RXSJ124830.1$-$594449 & 48 	& Diff.	 Spike?		\\
1RXSJ124849.0+333454 & 23 	& (no object)		\\
1RXSJ140559.3$-$411230 & 277 	& Diff. Spike		\\
1RXSJ144359.5+443124 & 42 	& (no object)		\\
1RXSJ153517.4$-$410958 & 295 	& Diff. Spike		\\
1RXSJ162609.7$-$242245 & 5 	& Neb./(no object)	\\
1RXSJ163910.7+565637  & 2 	& (no object)		\\
1RXSJ173253.6$-$371200 & 80 	& (no object)		\\
1RXSJ182102.0$-$161309 & 18 	& Neb./(no object)	\\
1RXSJ231117.9$-$094615 & 30 	& (no object)		\\

\enddata
\tablenotetext{a}{total number of  \usno\ objects in the
associated background fields}
\tablenotetext{b}{Neb.=nebulosity in field; Diff. Spike=stellar
diffraction spikes in field; (no object)=No obvious optical counterpart}
\tablecomments{Table contains information on fields which did not
contain any \usno\ objects, which were then visually inspected using
the DSS plate.  See Sec.~\ref{sec:blank}}
\end{deluxetable}

\begin{deluxetable}{lrr}
\scriptsize
\tablewidth{300pt}
\tablecaption{SIMBAD Source Types in the \pid$>$90\% Catalog \label{tab:sourcetypes}}
\tablehead{
\colhead{Class} & 
\colhead{N (\pid$>$90\% Cat.)} &
\colhead{N (\pid$>$50\% Cat.)} 
} 
\startdata
Algol		& 45 	& 61	\\
RS CVn		& 116	& 131	\\
W UMa		& 26	& 37	\\
T-Tauri		& 137	& 198	\\
Symbiotic Stars	& 2	& 2	\\
White Dwarfs	& 14 	& 57	\\
Dwarf Novae	& 9	& 33	\\
Cataclysm. Var. & 7	& 24 	\\
AGN		& 9	& 58	\\
Quasar		& 14	& 375	\\
Seyferts (1/2)	& 131	& 287	\\
BL Lac		& 7 	& 76	\\
Unclassified 	& 1362	& 4600	\\

\enddata
\end{deluxetable}

\begin{deluxetable}{lrrr|r}
\scriptsize
\tablewidth{300pt}
\tablecaption{Number of Identified Optical Counterparts \label{tab:number}}
\tablehead{
\colhead{\pid} & 
\colhead{N (single ID)\tablenotemark{a}} &
\colhead{N (binary ID)\tablenotemark{b}} &
\colhead{N (``blank'' fields)\tablenotemark{c}} &
\colhead{Totals} 
} 
\startdata
$\ge98$\%		& 2705		&	6	&	30	& 2741			\\
98$>$\pid$\ge$90\%	& 2787		&	25	& 	--	& 2812			\\
90$>$\pid$\ge$50\%	& 5809		&	441	&	--	& 6252			\\ \tableline
Total			& 11301		&	472	&	30	& 11803	\\

\enddata
\tablenotetext{a}{Sec.~\ref{sec:single}}
\tablenotetext{b}{Sec.~\ref{sec:binary}}
\tablenotetext{c}{Sec.~\ref{sec:blank}}
\end{deluxetable}

\end{document}


\appendix 
\begin{center}
{\bf Appendix: Tables of Cross-Identifications between the ROSAT/Bright Source Catalog and USNO-A2 Catalogs}\\
\end{center}

These four tables make up the appendix for the paper: ``XID:
Cross-Identifications of ROSAT/Bright Source Catalog Sources with
USNO-A2 Optical Point Sources''.  The production and meaning of all
values in this table are given in that work, and are briefly described
here:

\begin{table}[h]
\begin{tabular}{rcrr}
Table	&	Content			& From Sec. 	& \# Objects	\\ \hline
\ref{tab:98}	& \pid$\ge$98\% 		& 3.1		&2705 \\
\ref{tab:90}	& 98\%$>$\pid$\ge$90\% 		& 3.1		&2787	\\
\ref{tab:50} 	& 90\%$>$\pid$\ge$50\%		& 3.1		&5809	\\
\ref{tab:pair}	& Optical Pairs \pid$\ge$50\%	& 3.2		&472 	\\ 	\hline
\end{tabular}
\end{table}

The columns of each table (except Table~\ref{tab:pair}, see below) have the
following meanings:

\begin{table}[h]
\begin{tabular}{rll}
column & Heading	& Description \\ \hline 
1	& 1RXS	& The ROSAT/BSC Name for the X-ray source \\
2	& PSPC c/s ($\sigma$) & RASS/BSC Catalog X-ray countrate, and 1$\sigma$ uncertainty \\ 
3	& \pid	& calculated fractional probability of unique association between the X-ray and \usno\ objects \\
4	& $B_{\rm USNO A2}$ & the \usno\ $B$ magnitude \\
5	& USNO A2	& the \usno\ source name/position (hhmmss.ss+/-ddmmss.s) \\
6	& SIMBAD crossID & list of all objects $<$10\arcsec from the \usno\ position\\
7	& Type 	& SIMBAD object population	\\
8	& Class & Classification of the SIMBAD object \\
9	& $B$ : $V$ & SIMBAD $B$ and $V$ magnitudes \\
10	& Comments	& Comment code	 \\ \hline
\end{tabular}
\end{table}

\noindent In Table~\ref{tab:pair}, the magnitudes and \usno\
designations of the two sources are separated by a backslash. Please
be aware of the points of caution regarding the ``Optical Pairs''
counterparts.  In column 7, the source types are often abbreviated
according to the following scheme, which closely follows the source
types used by SIMBAD:
\clearpage
\begin{table}[h]
\begin{tabular}{ll|ll|ll} 
Abbrev. 	& Meaning		&Abbrev. 	& Meaning			&Abbrev. 	& Meaning		\\ \hline
$*$	& Star 				&HiPM*   & High Proper-motion Star        	&SN      & SuperNova                    \\ 
$**$	& Stellar binary 		&HMXB    & High Mass X-ray Binary         	&Spec. Bin. & Spectroscopic Binary       \\
$*$iC	& Star in Cluster 		&IR      & Infra-red object               	&Sy1     & Seyfert 1                     \\
$**$mul	& multiple stellar system 	&Glob. Clust.    & Globular Cluster       	&Sy2     & Seyfert 2                     \\
$*$Neb	& Star in Nebula		&Gal.    & Galaxy                         	&TT      & T-Tauri type star             \\
Clust.  & Cluster			&GiG     & Galaxy in Group of Galaxies    	&UV      & Ultra-violet emission source  \\
CV	& Cataclysmic Variable 		&LMXB    & Low Mass X-ray Binary          	&V*      & Variable Star                 \\
Ceph. 	& Cephied Variable 		&LSBG   & Low Surface Brightness Galaxy   	&WD      & White Dwarf                   \\
DN	& Dwarf Nova			&PN      & Planetary Nebula               	&WR *    & Wolf-Rayet Star               \\
Ecl. Bin. & Eclipsing Binary (and type)	&Rad.    & Radio source                   	&X       &  X-ray source                 \\
Em. *   & Emission line star            &Rot. Var. *& Rotationally Variable Star  	&YSO     & Young Stellar Object          \\ \hline
\end{tabular}
\end{table}

Additional lines are given for each SIMBAD object within 10\arcsec\ of
the USNO counterpart (20 \arcsec for the optical pairs) .  Note that
the counterpart for which the \pid\ applies is the \usno\ counterpart;
the SIMBAD identifications are listed to provide {\it possible}
identifications of this source. The SIMBAD $B$ and $V$ magnitudes
provide a point of comparison with the \usno\ $B$ magnitude which may
help in identifying the \usno\ object, however the \usno\ $B$
magnitudes are subject to certain systematic errors, and should be
viewed with caution.

Column 10 holds comment codes of which, at present, there is only one: P=SIMBAD
object has been previously identified as this RASS/BSC source.

\clearpage
\newpage
\texttt{}
\fontsize{6pt}{4pt}\selectfont

\setlongtables
\begin{longtable}{rrrrr|rrrrl}
\caption{Catalog of RASS/BSC -- \usno\ Associations, with \pid$\ge$98\%, and SIMBAD Sources within 10\arcsec \label{tab:98}} \\ \hline
1RXS & PSPC c/s ($\sigma$) & \pid & $B_{\rm USNO A2}$ & USNO A2 &
SIMBAD crossID & Type & Class & $B$ : $V$ & Comments \\ \hline
\endfirsthead
\multicolumn{10}{c}{Table~\ref{tab:98}:  \pid$\ge$98\% (cont.)} \\ \hline 
1RXS & PSPC c/s ($\sigma$) & \pid & $B_{\rm USNO A2}$ & USNO A2 &
SIMBAD crossID & Type & Class & $B$ : $V$ & Comments \\ \hline
\endhead
\hline \hfil continued... \\ 
\endfoot
\\ \hline 
\endlastfoot
\input tabA1.tex
\end{longtable}

\clearpage
\newpage

\begin{longtable}{rrrrr|rrrrl} 
\caption{Supplementary 90\%  RASS/BSC Associations in USNO Catalog, and
SIMBAD Sources ithin 10 arcsec \label{tab:90}} \\ \hline
1RXS & PSPC c/s ($\sigma$) & \pid & $B_{\rm USNO A2}$ & USNO A2 &
SIMBAD crossID & Type & Class & $B$ : $V$ & Comments \\ \hline
\endfirsthead
\multicolumn{10}{c}{Table~\ref{tab:90}:  Supplementary  90\%(cont.)} \\ \hline 
1RXS & PSPC c/s ($\sigma$) & \pid & $B_{\rm USNO A2}$ & USNO A2 &
SIMBAD crossID & Type & Class & $B$ : $V$ & Comments \\ \hline
\endhead
\hline \hfil continued... \\ 
\endfoot
\\ \hline 
\endlastfoot
\input tabA2.tex
\end{longtable}

\clearpage
\newpage

\begin{longtable}{rrrrr|rrrrl}
\caption{Supplementary 50\%  RASS/BSC Associations in USNO Catalog, and
SIMBAD Sources Within 10 arcsec \label{tab:50}} \\ \hline
1RXS & PSPC c/s ($\sigma$) & \pid & $B_{\rm USNO A2}$ & USNO A2 &
SIMBAD crossID & Type & Class & $B$ : $V$ & Comments \\ \hline 
\endfirsthead
\multicolumn{10}{c}{Table~\ref{tab:50}:  Supplementary 50\%  (cont.)} \\  \hline 
1RXS & PSPC c/s ($\sigma$) & \pid & $B_{\rm USNO A2}$ & USNO A2 &
SIMBAD crossID & Type & Class & $B$ : $V$ & Comments \\ \hline
\endhead
\hline \hfil continued... \\ 
\endfoot
\\ \hline 
\endlastfoot
\input  tabA3.tex
\end{longtable}

\clearpage
\newpage

\begin{longtable}{rrrrr|rrrrl}
\caption{RASS/BSC Cross IDs with Paried \usno\ objcts \label{tab:pair}} \\ \hline
1RXS & PSPC c/s ($\sigma$) & \pid & $B_{\rm USNO A2}$ & USNO A2 &
SIMBAD crossID & Type & Class & $B$ : $V$ & Comments \\ \hline 
\endfirsthead
\multicolumn{10}{c}{Table~\ref{tab:pair}: Pair Associations, \pid$\ge$50\% (cont.)} \\  \hline 
1RXS & PSPC c/s ($\sigma$) & \pid & $B_{\rm USNO A2}$ & USNO A2 &
SIMBAD crossID & Type & Class & $B$ : $V$ & Comments \\ \hline
\endhead
\hline \hfil continued... \\ 
\endfoot
\\ \hline 
\endlastfoot
  1RXSJ000202.5$-$103030  &     0.0609 (0.0162)  &    0.596 &      16.9/15.4  &    000202.09$-$103022.8/000202.96$-$103037.9 &              \nodata &              \nodata &    \nodata &         \nodata &       \\ 
  1RXSJ000813.3$-$005752  &     0.0507 (0.0135)  &    0.698 &      15.8/17.7  &    000812.21$-$005756.2/000813.25$-$005753.3 &              \nodata &              \nodata &    \nodata &         \nodata &       \\ 
    1RXSJ001000.6+490655  &     0.0506 (0.0133)  &    0.559 &      16.5/16.0  &        001000.08+490638.9/001001.35+490659.0 &              \nodata &              \nodata &    \nodata &         \nodata &       \\ 
  1RXSJ001049.7$-$081717  &     0.1019 (0.0194)  &    0.808 &      13.6/16.3  &    001049.14$-$081651.6/001049.60$-$081722.2 &             PHL 2695 &          Blue object &        ./. &          18.5 : &       \\ 
  1RXSJ001248.7$-$084705  &     0.1232 (0.0214)  &    0.569 &      16.2/16.3  &    001247.92$-$084700.3/001248.60$-$084644.4 &              \nodata &              \nodata &    \nodata &         \nodata &       \\ 
  1RXSJ001454.6$-$155215  &     0.0535 (0.0151)  &    0.713 &      17.3/17.0  &    001454.84$-$155213.4/001454.97$-$155208.9 &              \nodata &              \nodata &    \nodata &         \nodata &       \\ 
    1RXSJ002158.0+491245  &     0.0737 (0.0142)  &    0.529 &      14.4/17.1  &        002156.87+491239.6/002158.17+491243.1 &              GJ 3030 &                    * &       M2.5 &               : &       \\ 
  1RXSJ002605.4$-$185456  &     0.1336 (0.0242)  &    0.685 &      15.0/15.0  &    002604.91$-$185453.0/002605.56$-$185454.2 &              \nodata &              \nodata &    \nodata &         \nodata &       \\
\end{longtable}